\documentclass[10pt,journal]{IEEEtran}
\DeclareMathSizes{10}{9}{6}{5}

\usepackage{tabularx}
\usepackage{amsmath}
\usepackage{amssymb}
\usepackage{graphicx}
\usepackage{subcaption}
\usepackage{float}
\usepackage{siunitx}
\usepackage{url}
\usepackage{algorithm}
\usepackage{algorithmic}
\usepackage{xcolor}
\usepackage{booktabs}
\usepackage{multirow}
\usepackage{cite}
\usepackage{ulem}
\usepackage{flushend}
\usepackage{textcomp}

\usepackage[colorlinks=true, citecolor=blue, linkcolor=blue, urlcolor=blue]{hyperref}

\begin{document}
\title{Optimal Microgrid Operation with Open-cycle Ocean Thermal Energy Conversion for Islands}

\author{Xiaoyu Fu~\IEEEmembership{Student Member,~IEEE}, Yang~Yang,~\IEEEmembership{Member,~IEEE}
and~Yonghua~Song,~\IEEEmembership{Fellow,~IEEE}
\thanks{The authors are with the State Key Laboratory of Internet of Things for Smart City, University of Macau, Macau, China (e-mail: yc67003@um.edu.mo; yangy@um.edu.mo).}
\vspace{-12pt}}
\markboth{Manuscript submitted for review}
{Shell \MakeLowercase{\textit{et al.}}: Bare Demo of IEEEtran.cls for IEEE Journals}

\maketitle
\begin{abstract}
Ocean thermal energy conversion (OTEC) is a zero-carbon technology that harnesses the ocean’s thermal gradient to generate electricity. 
Among OTEC variants, open-cycle OTEC is particularly attractive to island communities, as it can co-generate electricity and freshwater. 
This paper develops an integrated model that captures both the thermodynamic process of open-cycle OTEC and its operational role in an island microgrid. 
A two-stage robust scheduling model is formulated for the island microgrid, with a budget uncertainty set to capture the renewable output deviations. 
The resulting model is solved via an inexact column-and-constraint generation algorithm, which accelerates convergence by permitting inexact solutions of the first-stage problem in early iterations. 
Numerical experiments demonstrate that open-cycle OTEC can fully substitute for conventional generators on island microgrids and provide more reliable and dispatchable output than volatile renewable sources.
\end{abstract}


\IEEEpeerreviewmaketitle

\section{Introduction}
\subsection{Research Background}
The ever-increasing carbon emissions from fossil fuels have accelerated the global transition to renewable energy sources (RESs).
In this context, ocean thermal energy conversion (OTEC), which harnesses the temperature difference between warm surface seawater and cold deep seawater to generate electricity, offers a promising alternative to fossil fuels \cite{claude1930power}. 
Unlike intermittent RESs such as solar and wind, OTEC can deliver stable power generation, with an estimated global potential of 8 to 10 TW \cite{du2022growth}.

OTEC systems are mainly classified into two types based on the thermal cycle path: closed and open \cite{sanjivy2026harnessing}. 
In a closed-cycle OTEC system, a low-boiling-point fluid (e.g., ammonia) is vaporized by warm surface seawater to drive a turbine, then condensed by cold deep seawater and recirculated. 
In an open-cycle system, seawater itself serves as the working fluid.
The warm surface seawater is flash-evaporated under vacuum to produce steam, which then drives the turbine and is further condensed into freshwater by cold deep seawater.
This dual production of electricity and freshwater makes open-cycle OTEC particularly attractive to island communities \cite{penny1984open}.

Yet most existing OTEC studies emphasize performance improvement in closed-cycle systems through working fluid selection and device-level design, while the open cycle remains largely overlooked \cite{zhang2018review,langer2022bigger,chang2025numerical}.
This necessitates the investigation of open-cycle OTEC.
Furthermore, when deploying the open-cycle OTEC on an island, it must be integrated into a microgrid that balances electricity and water demand through coordinated control.
Within the integrated microgrid, although open-cycle OTEC is steadier than solar and wind, its output may still vary with seawater temperatures in different seasons.
This calls for in-depth research on optimization models of island microgrids with open-cycle OTEC to ensure secure operation under adverse conditions.
Driven by the above considerations, this paper investigates the modeling of open-cycle OTEC and its integration into island microgrids.

\subsection{Contributions}

The main contributions of this paper are summarized as:
\begin{enumerate}
    \item An open-cycle OTEC model is developed based on thermodynamic principles, consisting of a flash evaporator, a steam turbine, a condenser, seawater pumps, and an exhaust system.
    A novel operational framework is introduced where four distinct operation statuses are determined by low-temperature and unit commitment signals to reflect realistic conditions.
    \item A two-stage robust microgrid operation model incorporating OTEC units is formulated to find the most economical operation scheme.
    A budget uncertainty set is constructed to capture the forecast errors of wind and solar power.
    We further investigate the microgrid operation model to glean operational insights: 
    a) Low-temperature conditions in winter can lead to frequent low-efficiency operation and reduced OTEC output, nearly quadrupling total costs compared to those in other seasons;
    b) open-cycle OTEC has the potential to fully replace conventional generators on island microgrids and provide more reliable output than volatile RESs.
    \item The resulting operation model is a large-scale robust problem, and we exploit an inexact column-and-constraint generation (iCCG) algorithm to solve it efficiently.
    With extensive numerical studies, we show the best parameter configurations for the iCCG algorithm and when it can achieve faster convergence than the conventional column-and-constraint generation algorithm. 
\end{enumerate}

\subsection{Literature Review}
\subsubsection{Ocean Thermal Energy Conversion} 
The current literature mainly focuses on thermal cycle design \cite{peng2022theoretical,liu2012progress, aydin2014off,zhang2023theoretical,zhou2021evaluation}, parameter configuration \cite{bernardoni2019techno,giostri2021off,yang2014analysis,ma2023thermodynamic}, and power potential estimation \cite{rajagopalan2013estimates,langer2021plant,calvo2025ocean} of closed-cycle OTECs.
In terms of cycle design, the literature focuses on supplementing the basic closed-cycle OTEC with additional components, such as heat recovery~\cite{liu2012progress,peng2022theoretical} and solar thermal preheating \cite{aydin2014off} to improve the power output and thermal efficiency. 
There are also studies exploring the multi-generation of cooling and freshwater by integrating power cycles with refrigeration \cite{zhang2023theoretical} or desalination stages~\cite{zhou2021evaluation}.
Regarding the parameter configuration, some researchers develop simulation tools to identify optimal parameters, such as heat exchanger geometry, to achieve the best thermodynamic properties \cite{bernardoni2019techno} or minimum levelized cost of electricity \cite{giostri2021off}, while others evaluate the influence of different working fluids on the performance of closed-cycle OTEC systems~\cite{yang2014analysis,ma2023thermodynamic}.
Furthermore, the power potential of OTEC is investigated at both global and regional levels, considering climate change \cite{du2022growth} and horizontal transportation of seawater  \cite{rajagopalan2013estimates} to provide deployment recommendations for different regions~\cite{langer2021plant,calvo2025ocean}.
Our paper differs from the above studies, which mainly focus on the closed-cycle OTEC, by investigating the open-cycle OTEC instead, and exploring its integration within an island microgrid.

\subsubsection{Microgrid Operation Problem} 
Our study also contributes to the literature on microgrids, which can be characterized by two key features.
First, they involve multi-energy coupling such as electricity, heat, natural gas, and hydrogen~\cite{siqin2022two,fei2024two,cheng2026generalized}, introducing endogenous and operational complexity.
Second, the high penetration of RESs such as photovoltaic (PV) and wind turbine (WT) brings intermittency and stochasticity to the system \cite{roald2023power}.
In response, the literature develops stochastic and robust optimization methods to hedge against the uncertainties.
Stochastic optimization relies on scenario generation to capture the probability distribution of uncertain variables \cite{wang2012chance}.
Robust optimization defines an uncertainty set to ensure the system remains functioning under the worst-case scenario~\cite{liu2018economic}.
Both approaches are often embedded within a two-stage optimization framework and widely applied in microgrid operation problems, e.g., unit commitment~\cite{blanco2017efficient} and economic dispatch~\cite{zhang2024strategic}.
Following these frameworks, we develop a two-stage robust optimization model for microgrids integrating OTECs to satisfy both the electricity and freshwater demand of islands, while explicitly accounting for the RES uncertainties.

\subsubsection{Decomposition Algorithm}
Modeling diverse microgrid devices and accounting for uncertainties yields a large-scale, two-stage robust optimization problem.
Column-and-constraint generation (CCG) is a widely adopted decomposition algorithm to solve two-stage robust optimization problems \cite{zeng2013solving}.
The CCG algorithm introduces decision variables and constraints associated with worst-case scenarios into the candidate master problem, and iteratively solves the candidate master problem and subproblem until convergence.
However, the size of the master problem increases as the iterations progress, and the computational burden escalates accordingly \cite{Zeng_2025}.
Recently, Tsang et al. \cite{tsang2023inexact} proposed an iCCG algorithm to accelerate the iteration process by allowing inexact solutions of the master problem in early iterations and verifying optimality only when necessary.
In this paper, we leverage the aforementioned techniques to efficiently solve the robust island microgrid operation problem.

The rest of this paper is organized as follows.
Section~\ref{sec: OTEC} presents the detailed modeling of open-cycle OTEC.
Section~\ref{sec: model} formulates the two-stage robust optimization model for microgrid operation.
Section~\ref{sec: solution method} introduces the solution methodology based on the iCCG algorithm.
Section~\ref{sec: numerical} discusses the numerical experiments, and Section~\ref{sec: conclusion} concludes the paper.
We present the nomenclature in online Appendix~A.

\section{Modeling of Open-cycle OTEC} \label{sec: OTEC}
In this section, we mathematically model the production process of the open-cycle OTEC so that it can be incorporated into the operation model of island microgrids.

\subsection{Components of Open-cycle OTEC}
\begin{figure}[!ht]
    \centering
    \includegraphics[width=\linewidth]{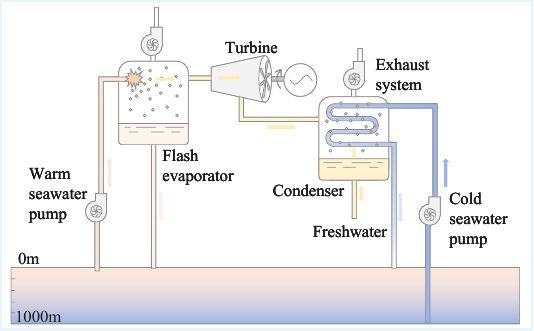}
    \caption{Open-cycle OTEC flow diagram}
    \label{fig:opencycle}
\end{figure}

The open-cycle OTEC generally consists of the following components: a flash evaporator, a steam turbine, a condenser, seawater pumps, and an exhaust system, as illustrated in Fig.~\ref{fig:opencycle}.
The warm seawater pump draws warm surface seawater and delivers it to the flash evaporator, where a portion of the warm seawater flashes into steam. 
The steam expands through the turbine to generate electricity, and is subsequently condensed into freshwater by the cold seawater in the condenser. 
Meanwhile, the exhaust system, consisting of several vacuum pumps, continuously removes the non-condensable gases to maintain the vacuum environment.
We introduce the detailed modeling of each component below.

\subsubsection{Flash Evaporator} In the first step, warm seawater is pumped to the flash evaporator.
The pressure inside the evaporator is lower than the saturation pressure of warm seawater at the inlet temperature, causing a portion of the seawater to flash into steam. 
During this process, the warm seawater releases a portion of its heat, and its temperature decreases.

According to the second law of thermodynamics, when the process reaches full thermodynamic equilibrium, the outlet warm seawater temperature $T^{\mathrm{ww,o}}_t$ would equal the steam saturation temperature $T^{\mathrm{evap}}$.
However, in practice, geometrical constraints and finite residence time will prevent the system from reaching perfect thermodynamic equilibrium~\cite{el2007flash}. 
Consequently, $T^{\mathrm{ww,o}}_t$ is higher than $T^{\mathrm{evap}}$, and the actual heat release falls short of the theoretical maximum.
Here, we introduce the effectiveness of the flash evaporator $\lambda^{\mathrm{evap}}_t$ as the ratio of actual heat release compared to the theoretical maximum heat release, which takes an empirical value of 0.95 under normal operating conditions \cite{sam1984experimental}.  
Let $T_t^{\mathrm{ww,i}}$ denote the temperature of inlet warm seawater, and $C^p$ denote the specific heat.
The actual heat transfer rate $q_t^{\mathrm{ww}}$ can be expressed as:
\begin{equation}
    q_t^{\mathrm{ww}}=\dot{m}_t^{\mathrm{ww}} C^p \lambda^{\mathrm{evap}}_t (T_t^{\mathrm{ww,i}} - T^{\mathrm{evap}}).
    \label{eq:otec1}
\end{equation}

The seawater flow rate $\dot{m}_t^{\mathrm{ww}}$ is constrained by its lower bound $\underline{\dot{m}}$ and upper bound $\overline{\dot{m}}$ due to pump capacity. 
To avoid large step changes in flow rate, which might induce pressure transients and cavitation risk, we introduce the ramping rate limit $\dot{m}^{\mathrm{ramp}}$:
\begin{align}
\underline{\dot{m}} &\leq \dot{m}_t^{\mathrm{ww}} \leq \overline{\dot{m}}, \label{eq:otec2} \\
-\dot{m}^{\mathrm{ramp}} &\le \dot{m}^{\mathrm{ww}}_{t} - \dot{m}^{\mathrm{ww}}_{t-1} \leq \dot{m}^{\mathrm{ramp}},\label{eq:otec3}
\end{align}
with the same constraints applied to the cold seawater flow $\dot{m}_t^{\mathrm{cw}}$.
Under steady-state conditions, the first law of thermodynamics implies that the heat released by warm seawater is utilized as the latent heat of vaporization $h^{\mathrm{evap}}_{\mathrm{fg}}$ to produce steam~\cite{el2007flash}. 
Therefore, the steam production rate $\dot{m}_t^{\mathrm{s}}$ is calculated as:
\begin{equation}
\dot{m}_t^{\mathrm{s}} = \frac{q_t^{\mathrm{ww}}}{h^{\mathrm{evap}}_{\mathrm{fg}}}.
\label{eq:otec4}
\end{equation}

\subsubsection{Turbine}
The vaporized steam then enters the turbine and expands isentropically, i.e., the entropy remains constant. 
This process causes a drop in pressure and temperature, converting thermal energy into mechanical work to rotate the shaft and generate power.
Since the vaporized steam is saturated, the specific enthalpy $h^{\mathrm{i}}$ and entropy $s^{\mathrm{i}}$ at the turbine inlet can be determined from the steam tables using the evaporation temperature $T^{\mathrm{evap}}$~\cite{wagner2008international}.
For example, at $T^{\mathrm{evap}}=21$ \si{\celsius}, the specific enthalpy and entropy for saturated steam are 2539.02~kJ/kg and 8.64 kJ/(kg$\cdot$K), respectively.
On the other hand, since the entropy remains constant, the specific entropy at the outlet of the turbine $s^{\mathrm{o}}$ equals $s^{\mathrm{i}}$. 
The specific enthalpy at the turbine outlet $h^{\mathrm{o}}$ can also be determined from the steam tables using $s^{\mathrm{o}}$ and the fixed condensation temperature $T^{\mathrm{cond}}=12$~\si{\celsius}.
Hence, the power output $w_{t}^{\mathrm{th}}$ is:
\begin{align}
w_{t}^{\mathrm{th}} = \dot{m}_t^{\mathrm{s}} (h^{\mathrm{i}} - h^{\mathrm{o}}).
\label{eq:otec5}
\end{align}
Considering the efficiencies $\eta_{\mathrm{turb}}$ and $\eta_{\mathrm{gen}}$ of the turbine and generator, the actual power generation is expressed as:
\begin{align}
w_t^{\mathrm{turb}} = w_{t}^{\mathrm{th}} \eta_{\mathrm{turb}} \eta_{\mathrm{gen}}.
\label{eq:otec6}
\end{align}

\subsubsection{Condenser}
The exhaust steam leaving the turbine is then isobarically condensed in the condenser, that is, the pressure remains constant in the condenser. 
During this process, the latent heat of the exhaust steam is released and transferred to the cold seawater. 

Given that the pressure drop from the turbine outlet to the condenser inlet is negligible, the specific enthalpy at the condenser inlet equals that at the turbine outlet $h^{\mathrm{o}}$.
Furthermore, assuming the steam exits the condenser as a saturated liquid, the outlet specific enthalpy is given by $h^{\mathrm{cond}}$.
Based on these premises, the latent heat released by the steam is~\cite{hernandez2022multi}:
\begin{equation}
q_t^{\mathrm{cw}} =\dot{m}_t^{\mathrm{s}} (h^{\mathrm{o}}-h^{\mathrm{cond}}). \label{eq:otec7}
\end{equation}
According to the first law of thermodynamics, the heat absorbed by the cold seawater is equal to the latent heat released during condensation.
Let $T^{\mathrm{cw,i}}_t$ and $T^{\mathrm{cw,o}}_t$ denote the inlet and outlet temperatures of the cold seawater, respectively.
The heat absorbed by the cold seawater $q_t^{\mathrm{cw}}$ is:
\begin{equation}
q_t^{\mathrm{cw}}=\dot{m}_t^{\mathrm{cw}}C^p(T^{\mathrm{cw,o}}_t - T^{\mathrm{cw,i}}_t). \label{eq:otec8}
\end{equation}
When there is sufficient cold water flow, the outlet cold seawater can be heated up to $T^{\mathrm{cond}} - \Delta T^{\mathrm{term}}$, where $T^{\mathrm{cond}}=12$~\si{\celsius} refers to the condensation temperature and $\Delta T^{\mathrm{term}}$ refers to the minimum allowable temperature difference in a heat exchanger. 
The difference reflects the thermodynamic driving force for heat transfer.
Suppose we further increase the cold seawater flow, the condensation capability will be higher. 
Nevertheless, this will lead to excessive pumping power and additional cost.
In contrast, an insufficient flow rate of cold seawater cannot fully condense the steam and may even disturb the normal operation of the OTEC system.
Therefore, to achieve the most economical operation, the optimal cold seawater flow should be controlled to ensure the outlet seawater temperature $T^{\mathrm{cw,o}}_t = T^{\mathrm{cond}} - \Delta T^{\mathrm{term}}$. 
In practice, the $\Delta T^{\mathrm{term}}$ can range from 2.8 \si{\celsius} to 5.6 \si{\celsius} \cite{guyer2018introduction}.

\subsubsection{Seawater Pumps} In both the evaporator and the condenser, seawater needs to be pumped to provide a heat source for evaporation and a cold source for condensation.
Given the warm seawater mass flow rate $\dot{m}_t^{\mathrm{ww}}$, the efficiency of seawater pumps $\eta
^{\mathrm{pump}}$, the gravitational acceleration $g:= 9.81~\mathrm{m/s^2}$, and the pump heads $\Delta h_t^{\mathrm{ww}}$ and $\Delta h_t^{\mathrm{cw}}$, the power consumed by the seawater pumps can be calculated as:
\begin{equation}
w_t^{\mathrm{ww}} = \frac{\dot{m}_t^{\mathrm{ww}} \Delta h_t^{\mathrm{ww}} g}{\eta^{\mathrm{pump}}}, \quad w_t^{\mathrm{cw}} = \frac{\dot{m}_t^{\mathrm{cw}} \Delta h_t^{\mathrm{cw}} g}{\eta^{\mathrm{pump}}}.\label{eq:otec9}
\end{equation}
Here, the pump heads $\Delta h_t^{\mathrm{ww}}$ and $\Delta h_t^{\mathrm{cw}}$ are defined as the energy per unit weight that must be supplied by the pumps to convey the fluid from the inlet to the outlet.
According to~\cite{zhou2021evaluation}, the pump head for warm seawater mainly involves the friction loss in the pipeline and the minor loss due to bending of the pipeline; whereas in the case of cold seawater, the head mainly involves friction loss, minor loss, and density variation, i.e., 
\begin{align}
  &\Delta h_t^{\mathrm{ww}} =\Delta h_t^{\mathrm{ww,fr}}+\Delta h_t^{\mathrm{ww,m}},\label{eq:otec10}\\
  &\Delta h_t^{\mathrm{cw}}=\Delta h_t^{\mathrm{cw,fr}}+\Delta h_t^{\mathrm{cw,m}}+\Delta h_t^{\mathrm{den}}\label{eq:otec11},
\end{align}
where $\Delta h_t^{\mathrm{ww,fr}}$ and $\Delta h_t^{\mathrm{cw,fr}}$ denote the friction heads of warm and cold seawater pumps, $\Delta h_t^{\mathrm{ww,m}}$ and $\Delta h_t^{\mathrm{cw,m}}$ denote the minor losses of warm and cold seawater, respectively; and $\Delta h_t^{\mathrm{den}}$ denotes the density head of the cold seawater pump that accounts for the pressure variation due to the density difference between the surface and the deep sea.
Let $L^{\mathrm{ww}}$ and $D^{\mathrm{ww}}$ denote the length and diameter of the warm seawater pipe, and $L^{\mathrm{cw}}$ and $D^{\mathrm{cw}}$ denote the length and diameter of the cold seawater pipe, respectively.
Let $v_t^{\mathrm{ww}}= \frac{4\dot{m}_t^{\mathrm{ww}}}{\rho\pi (D^{\mathrm{ww}})^2}$ and $v_t^{\mathrm{cw}}= \frac{4\dot{m}_t^{\mathrm{cw}}}{\rho\pi (D^{\mathrm{cw}})^2}$ represent the velocities of the seawater inside the pipes, and let $\rho$ denote the density of seawater. 
The head components can be approximated by the following empirical equations \cite{ma2023thermodynamic}:
\begin{align}
  &\Delta h_t^{\mathrm{ww,fr}}
  = 6.82 \times \frac{L^{\mathrm{ww}}}{(D^{\mathrm{ww}})^{1.17}} \times \left(\frac{v_t^{\mathrm{ww}}}{100}\right)^{1.85},\label{eq:otec12}\\
  &\Delta h_t^{\mathrm{cw,fr}}
  = 6.82 \times \frac{L^{\mathrm{cw}}}{(D^{\mathrm{cw}})^{1.17}} \times \left(\frac{v_t^{\mathrm{cw}}}{100}\right)^{1.85},\label{eq:otec13}\\
  &\Delta h_t^{\mathrm{ww,m}}
  = 60 \times \frac{({v_t^{\mathrm{ww}})}^{2}}{2g}, \quad \Delta h_t^{\mathrm{cw,m}}
  = 60 \times \frac{{(v_t^{\mathrm{cw}})}^{2}}{2g},\label{eq:otec14}\\
  &\Delta h_t^{\mathrm{den}} = L^{\mathrm{cw}} - \frac{\rho^{\mathrm{cw}}+\rho^{\mathrm{ww}}}{2 \rho^{\mathrm{cw}}}      L^{\mathrm{cw}}\label{eq:otec15}.
\end{align}
It is evident that the relationships \eqref{eq:otec12}-\eqref{eq:otec15} are nonlinear. 
To reduce the computational burden, we employ piecewise linearization techniques; see online Appendix B.

\subsubsection{Exhaust System} Gases such as oxygen and nitrogen are dissolved in seawater and will come out of solution in the deep vacuum environment of the cycle \cite{vega2013ocean}. 
Hence, an exhaust system consisting of vacuum pumps is equipped to remove the gases and maintain a vacuum environment.
For simplification, a fixed power consumption rate $\eta^{\mathrm{exh}}$ is introduced to calculate the exhaust system power consumption $w_t^{\mathrm{exh}}$: 
\begin{equation}
    w_t^{\mathrm{exh}} =  w^{\mathrm{th}}_{t}  \eta^{\mathrm{exh}}.
    \label{eq:otec16}
\end{equation}
According to \cite{penny1984open}, the power consumption rate of the exhaust system $\eta^{\mathrm{exh}}$ is approximately 7.6\%. 

In summary, the net power output can be expressed as:
\begin{align}
    w_t^{\mathrm{net}} = w_t^{\mathrm{turb}} - w_t^{\mathrm{ww}} - w_t^{\mathrm{cw}} - w_t^{\mathrm{exh}},
    \label{eq:otec17}
\end{align}
together with freshwater generated with a flow rate $\dot{m}_t^s$.

\subsection{Operational Modeling of Open-cycle OTEC}
To explore the conditions for the normal operation of each open-cycle OTEC unit in the microgrid $i\in \mathcal{N}^{\mathrm{OTEC}}$, we define the degree of superheat as the difference between inlet seawater temperature and evaporation temperature $\Delta T^{\mathrm{sup}}_{i,t}=T^{\mathrm{ww,i}}_{i,t}-T^{\mathrm{evap}}_i$.
At low inlet temperatures, warm seawater in the flash evaporator could hardly evaporate. 
According to~\cite{sam1984experimental}, the minimum degree of superheat for an evaporator's normal operation is 1.7~\si{\celsius}. 
We capture this physical trigger with a binary signal $x_{i,t}^{\mathrm{OTEC}}$, where $x_{i,t}^{\mathrm{OTEC}}=0$ indicates $\Delta T^{\mathrm{sup}}_{i,t} \ge 1.7$ \si{\celsius}, and $x_{i,t}^{\mathrm{OTEC}}=1$ otherwise.
Hence, for all $i \in \mathcal{N}^{\mathrm{OTEC}}, t \in \mathcal{T}$, we have:
\begin{align}
    -M_1 x_{i,t}^{\mathrm{OTEC}} \le \Delta T^{\mathrm{sup}}_{i,t}-1.7 \le M_1 (1-x_{i,t}^{\mathrm{OTEC}}),\label{eq:otec18}
\end{align}
where $M_1$ is a sufficiently large number.
Moreover, we define a binary variable $y_{i,t}^{\mathrm{OTEC}}$ to denote the unit commitment of OTECs, where $y_{i,t}^{\mathrm{OTEC}}=1$ indicates operation, and shutdown otherwise. 
Hence, four operation statuses are determined by $x_{i,t}^{\mathrm{OTEC}}$ and $y_{i,t}^{\mathrm{OTEC}}$:

\begin{enumerate}
  \item \textit{Normal shutdown}\;$(x_{i,t}^{\mathrm{OTEC}}=0,\ y_{i,t}^{\mathrm{OTEC}}=0)$: the unit is normally shut down. Pumps are off and deliver no seawater, and the OTEC neither generates electricity nor produces water.

  \item \textit{Normal operation}\;$(x_{i,t}^{\mathrm{OTEC}}=0,\ y_{i,t}^{\mathrm{OTEC}}=1)$: the unit operates normally. 
  Pumps consume power to deliver seawater, and the OTEC generates electricity and freshwater. 
  For the flash evaporation process, $\lambda^{\mathrm{evap}}_t$ takes the empirical value of 0.95 for application purposes \cite{sam1984experimental}.

  \item \textit{Low-temperature shutdown}\;$(x_{i,t}^{\mathrm{OTEC}}=1,\ y_{i,t}^{\mathrm{OTEC}}=0)$: in the shutdown state, the OTEC is unaffected by insufficient superheat. 
  The pumps are off, and no electricity or freshwater is produced.
  
  \item \textit{Low-temperature operation}\;$(x_{i,t}^{\mathrm{OTEC}}=1,\ y_{i,t}^{\mathrm{OTEC}}=1)$: the system runs in a standby mode. 
  The flash evaporator effectiveness $\lambda^{\mathrm{evap}}_{i,t}=0$ rather than $0.95$, meaning no heat will be released by warm seawater, and no steam will be generated.
  The OTEC neither generates electricity nor produces water due to the lack of steam,  and the pumps are forced to take off to save energy.
\end{enumerate}

The above four operation statuses can be translated to: 
\begin{align}
&\underline{\dot{m}}\,y_{i,t}^{\mathrm{OTEC}}(1-x_{i,t}^{\mathrm{OTEC}})\le \dot{m}_{i,t}^{\mathrm{ww}}\le\overline{\dot{m}}\,y_{i,t}^{\mathrm{OTEC}}(1-x_{i,t}^{\mathrm{OTEC}}),\label{eq:otec19}\\
&\underline{\dot{m}}\,y_{i,t}^{\mathrm{OTEC}}(1-x_{i,t}^{\mathrm{OTEC}})\le\dot{m}_{i,t}^{\mathrm{cw}}\le\overline{\dot{m}}\,y_{i,t}^{\mathrm{OTEC}}(1-x_{i,t}^{\mathrm{OTEC}}),\label{eq:otec20}\\
&\lambda^{\mathrm{evap}}_{i,t}=0.95\,(1-x_{i,t}^{\mathrm{OTEC}}).\label{eq:otec21}
\end{align}
Note that \eqref{eq:otec19} and \eqref{eq:otec20} contain products of two binary variables, namely $y_{i,t}^{\mathrm{OTEC}}(1-x_{i,t}^{\mathrm{OTEC}})$. We decouple this nonlinear term using linearization techniques; see online Appendix C.

In summary, by controlling the flow of inlet warm seawater and cold seawater under different temperatures, we are able to control the power output and the freshwater production rate.  
The decision variables of the open-cycle OTEC, including the operation status $y_{i,t}^{\mathrm{OTEC}}, x_{i,t}^{\mathrm{OTEC}}$, net power output $w_{i,t}^{\mathrm{net}}$, pump power consumption $w_{i,t}^\mathrm{ww}, w_{i,t}^\mathrm{cw}$, exhaust system power consumption $w_{i,t}^\mathrm{exh}$ and water production rate $\dot{m}_{i,t}^{\mathrm{s}}$, are subject to the feasibility set
$\Phi^{\mathrm{OTEC}}(T_{i,t}^{\mathrm{ww,i}},T_{i,t}^{\mathrm{cw,i}}) := \{\eqref{eq:otec1}-\eqref{eq:otec21}\}.$

\section{Microgrid Operation Model} \label{sec: model}
In this paper, we consider a microgrid deployed on a remote island without external grid interconnection.
Open-cycle OTEC plants are deployed to supply both electricity and freshwater. 
The electricity production is complemented by RESs, including PV arrays and WTs, as well as a fleet of conventional generators.
Energy storage (ES) units and a water sink are deployed to balance the supply and demand of electricity and freshwater. 
To address the intermittency of RESs without relying on the probability distributions that are generally unavailable in practice, we formulate the microgrid operation model as a two-stage robust optimization problem.

\subsection{First-Stage Commitment}
The first stage represents the day-ahead \textit{here-and-now} decisions, comprising a set of commitments determined for each time period $t \in \mathcal{T}$ prior to the realization of uncertainties.
Decisions include the operational states of generators and OTEC units, together with the scheduled output and spinning reserve of generators.

\subsubsection{Generator Commitment}
For each generator $i \in \mathcal{N}^{\mathrm{gen}}$, first-stage variables include the binary commitment state $y^{\mathrm{gen}}_{i,t}$, start-up action $z^{\mathrm{gen}}_{\mathrm{on},i,t}$ and shut-down action $z^{\mathrm{gen}}_{\mathrm{off},i,t}$.
Because it takes several hours for generators to complete state transitions, the minimum uptime $\overline{T}_i$ and downtime $\underline{T}_i$ must be enforced. 
\begin{align}
& y^{\mathrm{gen}}_{i,t}-y^{\mathrm{gen}}_{i,t-1}
   = z^{\mathrm{gen}}_{\mathrm{on},i,t}-z^{\mathrm{gen}}_{\mathrm{off},i,t}, \qquad \forall t \in \mathcal{T},
   \label{mgt:logic2} \\
& z^{\mathrm{gen}}_{\mathrm{on},i,t}+z^{\mathrm{gen}}_{\mathrm{off},i,t}\le 1, \qquad \qquad \qquad ~~  \forall t \in \mathcal{T}, \label{mgt:mutex} \\
& \sum_{\tau=t}^{t+\overline{T}_i-1}
  y^{\mathrm{gen}}_{i,\tau}\ge
 z^{\mathrm{gen}}_{\mathrm{on},i,t}\,\overline{T}_i,   \qquad \forall t \in \{1,\dots,|\mathcal{T}|-\overline{T}_i+1\} ,\label{mgt:mindown} \\
& \sum_{\tau=t}^{t+\underline{T}_i-1}
  \bigl(1-y^{\mathrm{gen}}_{i,\tau}\bigr)\ge
  z^{\mathrm{gen}}_{\mathrm{off},i,t}\underline{T}_i,  \forall t \in \{1,\dots,|\mathcal{T}|-\underline{T}_i+1\}, \label{mgt:minup}
\end{align}
where we set $y^{\mathrm{gen}}_{i,0}:= 1$ for simplicity. 
Unlike generators, we do not consider start-up and shut-down constraints for OTEC.
This is because OTEC components can generally respond to operational changes within one hour \cite{hata2023introduction,bharathan1990conceptual}.

To accommodate upcoming fluctuations in renewable generation and ensure demand satisfaction, the scheduled power $p^{\mathrm{sch}}_{i,t}$ and the spinning reserve $r_{i,t}^{\mathrm{U}}$ and $r_{i,t}^{\mathrm{D}}$ of generator $i \in \mathcal{N}^{\mathrm{gen}}$ shall be determined in advance, limited by the respective minimum and maximum power outputs $\underline{p}_i^{\mathrm{gen}}, \overline{p}_i^{\mathrm{gen}}$.
Moreover, ramping constraints denoted by $p_i^{\mathrm{ramp}}$ are introduced to manage the physical limitations of generation units:
\begin{align}
  & \underline{p}_i^{\mathrm{gen}}\,y^{\mathrm{gen}}_{i,t}\le p^{\mathrm{sch}}_{i,t} - r_{i,t}^{\mathrm{D}},  && \forall t \in \mathcal{T}, \label{mgt:schmin}\\
  & p^{\mathrm{sch}}_{i,t} + r_{i,t}^{\mathrm{U}}
  \le \overline{p}_i^{\mathrm{gen}}y^{\mathrm{gen}}_{i,t},  && \forall t \in \mathcal{T}, \label{mgt:schmax}\\
  & -p_i^{\mathrm{ramp}} \le p^{\mathrm{sch}}_{i,t}-p^{\mathrm{sch}}_{i,t-1}
       \le p_i^{\mathrm{ramp}},  && \forall t \in \mathcal{T}. \label{mgt:ramp}
\end{align}

\subsection{Modeling of Uncertainty Set}
The second-stage \textit{wait-and-see} decisions are made after the realization of uncertainties, including the fluctuations in PV and WT power outputs.
Instead of enumerating all scenarios, we construct a budget uncertainty set $\mathcal{U}$ following \cite{zhang2024two}. 
Let $\mathcal{N}^{\mathrm{RES}} = \mathcal{N}^{\mathrm{PV}} \cup \mathcal{N}^{\mathrm{WT}}$ denote the set of all RESs,
and let binary variables $\epsilon_{i,t}^{\pm}$ indicate whether the output of unit $i$ deviates to its upper or lower bound.
Furthermore, a budget parameter $\Gamma$ is introduced to limit the total number of time periods where such deviations exist.
The uncertainty set $\mathcal{U}$ is defined as:
\begin{equation}
\mathcal{U} = \left\{ \epsilon_{i,t}^{\pm} \left| 
\begin{aligned}
& \sum_{i \in \mathcal{N}^{\mathrm{RES}}}\sum_{t \in \mathcal{T}} (\epsilon_{i,t}^{+} + \epsilon_{i,t}^{-}) \le \Gamma, \\
& \epsilon_{i,t}^{+} + \epsilon_{i,t}^{-} \le 1, \qquad \forall i \in \mathcal{N}^{\mathrm{RES}}, t \in \mathcal{T}\\ 
&\epsilon_{i,t}^{\pm} \in \{0,1\}, ~~\qquad \forall i \in \mathcal{N}^{\mathrm{RES}}, t \in \mathcal{T}
\end{aligned}
\right. \right\}.\label{uncertain-cons}
\end{equation}
Let $\delta_i$ denote the maximum relative deviation of each unit, and let $\bar{p}_{i,t}$ represent the deterministic day-ahead forecast value at time $t$.
The power output $\tilde{p}_{i,t}$ under a certain scenario can be determined by the realization of $\epsilon_{i,t}^{\pm}$:
\begin{equation}
\tilde{p}_{i,t} = \bar{p}_{i,t} \left( 1 + \delta_i (\epsilon_{i,t}^+ - \epsilon_{i,t}^-) \right), \quad \forall i \in \mathcal{N}^{\mathrm{RES}}, t \in \mathcal{T}.
\end{equation}

\subsection{Second-stage Dispatch}
Second-stage dispatch decisions are made after the uncertainty realizations are observed. 
These decisions enable rapid adjustments to meet the power and water balance requirements, including power schedules, water flow rates, and storage levels for each device and each time period $t \in \mathcal{T}$.

\subsubsection{OTEC Dispatch}
For each OTEC unit $i \in \mathcal{N}^{\mathrm{OTEC}}$, the second-stage dispatch variables include pump power consumption $w_{i,t}^{\mathrm{ww}}, w_{i,t}^{\mathrm{cw}}$, exhaust system power consumption $w_{i,t}^{\mathrm{exh}}$, net power generation $w_{i,t}^{\mathrm{net}}$, and freshwater production $\dot{m}_{i,t}^{\mathrm{s}}$.
These variables, constrained by \eqref{eq:otec1}--\eqref{eq:otec17}, represent the key interfaces between the OTEC subsystem and the broader microgrid.

\subsubsection{Generator Dispatch}
For each generator unit $i \in \mathcal{N}^{\mathrm{gen}}$, the actual output $p^{\mathrm{gen}}_{i,t}$ is decided to ensure operational feasibility and economic efficiency. 
As the scheduled output $p^{\mathrm{sch}}_{i,t}$ and the spinning reserve $r_{i,t}^\mathrm{U}$ and $r_{i,t}^\mathrm{D}$ have already been decided in the first stage, the actual output is restricted by:
\begin{align}
   p^{\mathrm{sch}}_{i,t} - r_{i,t}^{\mathrm{D}} \le p^{\mathrm{gen}}_{i,t}    \le p^{\mathrm{sch}}_{i,t}+r_{i,t}^{\mathrm{U}}, \qquad \forall t \in \mathcal{T}. \label{mgt:gen}
\end{align}

\subsubsection{Energy Storage Dispatch}
For each energy storage unit $i\in \mathcal{N}^{\mathrm{ES}}$, the hourly charging power $p^{\mathrm{ch}}_{i,t}$ and discharging power $p^{\mathrm{dis}}_{i,t}$ are dispatched in the second stage, constrained by 
the maximum charging and discharging rate $\overline{p}^{\mathrm{ES}}$.
Accordingly, the state of charge (SOC) $E_{i,t}$ evolves dynamically based on the previous state and current power schedules.
Let $\eta_i^{\mathrm{ch}}, \eta_i^{\mathrm{dis}}$ denote the charging and discharging efficiencies, and $\underline{E}_i,\overline{E}_i$ denote the lower and upper capacity limits of storage unit $i$, respectively.
The SOC is formulated as:
\begin{align}
& 0 \le p^{\mathrm{ch}}_{i,t} \le \overline{p}^{\mathrm{ES}}, ~~0 \le p^{\mathrm{dis}}_{i,t} \le  \overline{p}^{\mathrm{ES}}, && \forall t \in \mathcal{T}, \label{es:ch} \\
& E_{i,t} = E_{i,t-1}
     + \eta_i^{\mathrm{ch}}\, p^{\mathrm{ch}}_{i,t}
     - \frac{1}{\eta_i^{\mathrm{dis}}}\, p^{\mathrm{dis}}_{i,t}, && \forall t \in \mathcal{T},
  \label{es:soc} \\
& \underline{E}_i \le E_{i,t} \le \overline{E}_i, && \forall t \in \mathcal{T},
  \label{es:bounds}
\end{align}
where we set the initial SOC $E_{i,0}:= 250$ kW.

\subsubsection{Power Flow Dispatch}
In the second-stage dispatch, power flow constraints are incorporated to ensure network deliverability and track real-time power flows across all nodes $m \in \mathcal{N}^{\mathrm{bus}}$ and lines $l \in \mathcal{N}^{\mathrm{line}}$. 
Let $p_{m,t}$ denote the net power injection at bus $m$, $\overline{p}_{l}$ denote the transmission limit of network line $l$, and $\mathrm{PTDF}_{:,:}$ denote the entry of the power transfer distribution factor.
The power flow model is described as:
\begin{align}
& p_{m,t}
= \sum_{i\in \mathcal{N}^{\mathrm{RES}}_m} \tilde{p}_{i,t}
+ \sum_{i\in \mathcal{N}^{\mathrm{gen}}_m} p^{\mathrm{gen}}_{i,t}
+ \sum_{i\in \mathcal{N}^{\mathrm{OTEC}}_m} w_{i,t}^{\mathrm{net}}\notag\\
&\quad \quad+ \sum_{i\in \mathcal{N}^{\mathrm{ES}}_m} (p^{\mathrm{dis}}_{i,t}
- p^{\mathrm{ch}}_{i,t})-\mathrm{Ld}_{m,t}^{\mathrm{e}}+p^{\mathrm{dr,e}}_{m,t}, ~~\forall t \in \mathcal{T},
\label{eq:power_balance} \\
&\sum_{m\in \mathcal{N}^{\mathrm{bus}}}p_{m,t}=0, \qquad \qquad   \qquad \qquad \qquad ~~ \forall t \in \mathcal{T},\\
& -\overline{p}_{l} \leq
\sum_{m\in \mathcal{N}^{\mathrm{bus}}} \mathrm{PTDF}_{l,m}\,p_{m,t}
\leq \overline{p}_{l}, \quad \qquad ~~  \forall t \in \mathcal{T},
\label{eq:line_limits}
\end{align}
where $\mathcal{N}^{\mathrm{RES}}_m, \mathcal{N}^{\mathrm{gen}}_m, \mathcal{N}^{\mathrm{OTEC}}_m, \mathcal{N}^{\mathrm{ES}}_m$ denote the sets of RES, generator, OTEC and energy storage units connected to bus $m$, respectively, and $p^{\mathrm{dr,e}}_{m,t}$ represents the electric demand response at bus $m$, bounded above by the electricity load $\mathrm{Ld}_{m,t} ^{\mathrm{e}}$.

\subsubsection{Water Flow Dispatch}
We assume all OTEC units are connected to a shared water sink, which serves as a freshwater buffer to mitigate temporal mismatches between production and demand. 
One may also consider the case of multiple water sinks, albeit with a slightly more complex model.
The freshwater output $\dot{m}^{\mathrm{s}}_{i,t}$ from each OTEC unit $i\in \mathcal{N}^{\mathrm{OTEC}}$ is allocated between the sink $\dot{m}^{\mathrm{s\text{-}sink}}_{i,t}$ and direct load supply $\dot{m}^{\mathrm{s\text{-}ld}}_{i,t}$. 
Meanwhile, the sink releases water $\dot{m}^{\mathrm{sink\text{-}ld}}_{t}$ to compensate for any demand shortfall whenever needed.
Similar to SOC in energy storage dispatch, water level $h^{\mathrm{sink}}_{t}$ reflects the real-time storage status of the sink.
Let $\rho_{\mathrm{w}}$ denote the density of freshwater, $\texttt{A}$ denote the bottom area of the sink, and $\underline{h},\overline{h}$ denote the lower and upper limits of the water level, respectively.
The water sink dispatch model is formulated as:
\begin{align}
& \dot{m}^{\mathrm{s}}_{i,t} = \dot{m}^{\mathrm{s\text{-}ld}}_{i,t} + \dot{m}^{\mathrm{s\text{-}sink}}_{i,t}, \quad \forall t \in \mathcal{T},
\label{sink:split}\\
& h^{\mathrm{sink}}_{t} = h^{\mathrm{sink}}_{t-1} 
+ \left[ \sum_{i \in \mathcal{N}^{\mathrm{OTEC}}} \dot{m}^{\mathrm{s\text{-}sink}}_{i,t} 
- \dot{m}^{\mathrm{sink\text{-}ld}}_{t} \right] \frac{\Delta t}{\rho_{\mathrm{w}} \texttt{A}},  ~~ \forall t \in \mathcal{T}, 
\label{sink:level2} \\
& \underline{h} \le h^{\mathrm{sink}}_{t} \le \overline{h},\quad  \forall t \in \mathcal{T},
\label{sink:bounds}
\end{align}
where we set the initial water level $ h^{\mathrm{sink}}_{0}:= 3$ m.

Since OTEC freshwater production and sink drainage may not fully cover peak demand, we incorporate system-wide aggregated demand response for water consumption $\dot{m}^{\mathrm{dr,w}}_{t}$ upper bounded by the water demand $\mathrm{Ld}_t^{\mathrm{w}}$. 
Water balance is then expressed as:
\begin{align}
& \sum_{i \in \mathcal{N}^{\mathrm{OTEC}}} \dot{m}^{\mathrm{s\text{-}ld}}_{i,t} + \dot{m}^{\mathrm{sink\text{-}ld}}_{t} 
= \mathrm{Ld}_{t}^{\mathrm{w}}-\dot{m}^{\mathrm{dr,w}}_{t},&& \forall t \in \mathcal{T}.
\label{sink:demand}
\end{align}

\subsection{Objective Function}
The objective function aims to minimize the overall operational cost of the system under the worst-case realization of uncertainties. 
The first-stage commitment cost encompasses generator schedule and reserve costs, as well as generator commitment cost.
Regarding the second-stage cost, we consider the levelized operational expenses of OTECs, 
energy storage operating cost, and demand response cost paid to users for both electricity and water load adjustments.
Overall, the objective function can be expressed as:
\begin{align}
\min \big\{ &\sum_{t\in \mathcal{T}} \sum_{i\in \mathcal{N}^{\mathrm{gen}}}   \bigl(c^{\mathrm{gen}}_{\mathrm{on},i}z^{\mathrm{gen}}_{\mathrm{on},i,t} + c^{\mathrm{gen}}_{\mathrm{off},i}z^{\mathrm{gen}}_{\mathrm{off},i,t}+ c^{\mathrm{gen,s}}_i P^{\mathrm{sch}}_{i,t} + c^{\mathrm{gen,r}}_i r_{i,t}^{\mathrm{U}} \notag\\
&+ c^{\mathrm{gen,r}}_i r_{i,t}^{\mathrm{D}} \bigr)+ \max_{\mathbf{u} \in \mathcal{U}} \min \big[\sum_{t\in \mathcal{T}}\big( \sum_{i\in \mathcal{N}^{\mathrm{ES}}} c_i^{\mathrm{ES}}(p^{\mathrm{ch}}_{i,t}+ p^{\mathrm{dis}}_{i,t}) \notag\\
   & +\sum_{i\in \mathcal{N}^{\mathrm{OTEC}}}c_i^{\mathrm{OTEC}}(w_{i,t}^{\mathrm{exh}} +w_{i,t}^{\mathrm{ww}}  +w_{i,t}^{\mathrm{cw}})\notag\\
   &+ \sum_{i\in \mathcal{\mathcal{N}^{\mathrm{bus}}}} c_i^{\mathrm{dr,e}}p^{\mathrm{dr,e}}_{m,t} + c^{\mathrm{dr,w}} \dot{m}^{\mathrm{dr,w}}_{t} \big) \big] \big\}\label{obj:overall},
\end{align}
where $c^{\mathrm{gen}}_{\mathrm{on},i}, c^{\mathrm{gen}}_{\mathrm{off},i}, c^{\mathrm{gen,s}}_i, c^{\mathrm{gen,r}}_i, c^{\mathrm{ES}}_i, c^{\mathrm{exh}}_i, c^{\mathrm{OTEC}}_i, c^{\mathrm{dr,e}}_i, c^{\mathrm{dr,w}}$ denote the unit costs for generator start-up, generator shut-down, generator scheduled output, spinning reserve, energy storage operation, OTEC, electrical demand response, and water demand response, respectively.

\section{Solution Methodology} \label{sec: solution method}
\subsection{Model Reformulation}
To enable the application of efficient solution algorithms, the proposed model is first reformulated into a compact matrix form. Let $\mathbf{v}$ denote the first-stage decision variables, $\mathbf{u} \in \mathcal{U}$ denote the uncertain variables, and $\mathbf{w}$ denote the second-stage recourse variables, with detailed definitions provided in the online Appendix D.
The compact form is expressed as:
\begin{align}
\min_{\mathbf{v}} \quad & \mathbf{c}^\top \mathbf{v} +\max_{\mathbf{u} \in \mathcal{U}} \min_{\mathbf{w}} \mathbf{d}^\top \mathbf{w} \label{eq:compact_obj} \\
\text{s.t.} \quad & \mathbf{A} \mathbf{v} \geq \mathbf{b}  ,\label{eq:compact_first_ineq} \\
& \mathbf{D} \mathbf{w} \geq \mathbf{f} - \mathbf{G} \mathbf{v} - \mathbf{H} \mathbf{u}\;\rightarrow\; \boldsymbol{\pi}, \label{eq:compact_second_ineq}
\end{align}
where $\mathbf{c}$ and $\mathbf{d}$ are cost coefficients in the objective function~\eqref{obj:overall}.
The first-stage constraint matrix $\mathbf{A}$ and right-hand-side vector $\mathbf{b}$ correspond to constraints \eqref{eq:otec18}--\eqref{mgt:ramp}.
The second-stage coefficient matrices $\mathbf{D},\mathbf{G}$ and $\mathbf{H}$ and right-hand-side vector $\mathbf{f}$ correspond to constraints \eqref{eq:otec1}--\eqref{eq:otec17}, \eqref{uncertain-cons}--\eqref{sink:demand}, and $\boldsymbol{\pi}$ denotes the dual variables associated with the second-stage constraints~\eqref{eq:compact_second_ineq}.

The compact form has a max--min structure and cannot be solved directly. 
However, note that given $\mathbf{v}$ and $\mathbf{u}$, the second-stage subproblem is a linear program. 
By strong duality, the inner minimization can be replaced by its dual maximization, yielding an equivalent single-level maximization problem  $\max_{\boldsymbol{\pi},\mathbf{u}}
\left\{
\boldsymbol{\pi}^\top\left(\mathbf{f}-\mathbf{G}\mathbf{v}-\mathbf{H}\mathbf{u}\right)
\ \text{s.t.}\ 
\mathbf{D}^\top\boldsymbol{\pi}\le \mathbf{d}
\right\}$.
The reformulated objective contains a bilinear term $\boldsymbol{\pi}^\top \mathbf{H}\mathbf{u}$ coupling the continuous dual variables $\boldsymbol{\pi}$ with the binary uncertainty variables $\mathbf{u}$. 
To linearize this term, we denote $\mathbf{h}_k$ as the $k$-th column of $\mathbf{H}$ for $k \in [K]$, introduce an auxiliary variable vector $\mathbf{q}=\{q_k|q_k = u_k(\boldsymbol{\pi}^\top \mathbf{h}_k), \forall k \in [K]\}$, and use the Big-$M$ method to capture the bilinear relation.
Consequently, the second-stage problem is reformulated into a mixed integer linear program:
\begin{equation}\label{eq:sp} \tag{SP}
\begin{aligned}
Q(\mathbf{v}) =\max_{\mathbf{u}, \boldsymbol{\pi}, \mathbf{q}} \quad 
&\boldsymbol{\pi}^\top (\mathbf{f} - \mathbf{G} \mathbf{v}) - \sum_{k \in [K]} q_k \\
\text{s.t.} \quad 
& \mathbf{D}^\top \boldsymbol{\pi} = \mathbf{d}, \\
& \mathbf{0} \le \boldsymbol{\pi},\mathbf{u} \in \mathcal{U}, \\
& -M u_k \le q_k \le M u_k, && \forall k \in [K], \\
& q_k \le \boldsymbol{\pi}^\top \mathbf{h}_k + M(1-u_k), && \forall k \in [K], \\
& q_k \ge \boldsymbol{\pi}^\top \mathbf{h}_k - M(1-u_k), && \forall k \in [K].
\end{aligned}
\end{equation}

\subsection{Solution Algorithm}
As a popular approach in two-stage robust optimization problems, CCG iteratively solves the first-stage master problem and the second-stage subproblem, and augments the master problem with variables and worst-case constraints, progressively converging to the optimal solution.
However, in the proposed microgrid operation model, the master problem involves numerous binary variables and becomes increasingly difficult to solve as more worst-case cuts are added.
To mitigate this issue, we adopt the iCCG algorithm~\cite{tsang2023inexact} and briefly introduce it here to make the paper self-contained.

Unlike CCG, the iCCG algorithm does not require solving the master problem to global optimality in every iteration. 
Instead, it reduces the computational effort through \textit{exploration} and \textit{exploitation}.
In the \textit{exploration} phase, the master problem is solved inexactly with a relaxed tolerance and a dynamic lower bound.
This avoids the computational cost of proving optimality for a feasible solution and may accelerate the lower bound improvement.
In the \textit{exploitation} phase, the lower bound is set to a valid lower bound, forcing the solver to search more thoroughly to ensure convergence.

At iteration $j$, the master problem is:
\begin{equation}\label{eq:mp} \tag{MP}
\begin{aligned}
\min_{\mathbf{v}, \eta, \{\mathbf{w}_i\}_{i=1}^j}  & \mathbf{c}^\top \mathbf{v} + \eta \\
\text{s.t.} \quad & \mathbf{A} \mathbf{v} \ge \mathbf{b},  \\
& \eta \ge \mathbf{d}^\top \mathbf{w}_i, && \forall i = 1, \dots, j,  \\
& \mathbf{D} \mathbf{w}_i \ge \mathbf{f} - \mathbf{G} \mathbf{v} - \mathbf{H} \mathbf{u}^*_i, &&\forall i = 1, \dots, j,  \\
& \mathbf{c}^\top \mathbf{v} + \eta \ge \overline{L}, 
\end{aligned}
\end{equation}
where $\eta$ denotes the auxiliary variable approximating the second-stage recourse cost $Q(\mathbf{v})$, and $\mathbf{w}_i$ denotes the second-stage recourse variables associated with the worst-case scenario $\mathbf{u}^*_i$ at iteration $i$.
The dynamic lower bound $\overline{L}$ facilitates the solver to prune nodes with objective values lower than $\overline{L}$ in the branch and bound tree, thereby reducing the search efforts.

The iCCG algorithm is summarized in \textbf{Algorithm} \ref{alg:iccg}.
At each iteration $j$, the \eqref{eq:mp} is solved to obtain a candidate solution $\mathbf{v}_j^*$, along with objective $O_j$ and a lower bound  $L_j$ given by the solver. 
If $L_j$ is greater than $\bar{L}$, it is deemed a valid improvement.
In this case, we let $\ell \leftarrow j$ to record the current valid lower bound.
Subsequently, the \eqref{eq:sp} is solved with $\mathbf{v}_j^*$ to identify the worst-case scenario $\mathbf{u}_j^*$ and update the upper bound $\overline{U}$.
Note that any objective value associated with a candidate solution $\mathbf{v}_j^*$ represents a global upper bound of the optimal objective value for minimization problems.
If the inexact relative gap $(\overline{U}-O_j)/\overline{U}$ is larger than a tolerance $\tilde{\varepsilon}$, the algorithm proceeds with the \textit{exploration} phase.
The master problem is augmented with new recourse variables and cuts under the worst-case scenario $\mathbf{u}_j^*$, and the dynamic lower bound $\overline{L}$ is aggressively updated to $O_j$, which accelerates computation by solving inexactly.
Conversely, if the inexact relative gap falls below $\tilde{\varepsilon}$, indicating only a small improvement in $O_j$, the algorithm enters the \textit{exploitation} phase.
The dynamic lower bound $\overline{L}$ is reset to the safe anchor $L_\ell$ and the optimality tolerance for the master problem $\varepsilon^{\mathrm{MP}}=\{\varepsilon^{\mathrm{MP}}_{j'},j'\in \mathcal{J}\}$ is tightened by a contraction factor $\alpha$, where $\mathcal{J}$ is the set of iteration counts.

\begin{algorithm}[htbp]
\caption{iCCG Algorithm}
\label{alg:iccg}
\begin{algorithmic}[1]
\STATE \textbf{Initialization:} Set $\overline{L} \leftarrow 0$, $\overline{U} \leftarrow \infty$, $\ell \leftarrow 0$, $j \leftarrow 1$. Set tolerances $\varepsilon, \tilde{\varepsilon}$ and $\varepsilon^{\mathrm{MP}}$. Initialize scenario set $\mathcal{O} \leftarrow \{ \mathbf{u}^0 \}$.

\STATE \textbf{Step 1: Master Problem}
\STATE Solve \eqref{eq:mp} within relative gap $\varepsilon^{\mathrm{MP}}_j$. Get solution $(\mathbf{v}_j^*, \eta_j^*)$, objective $O_j$ and lower bound $L_j$.
\STATE Record valid lower bound. If $L_j > \bar{L}$, set $\ell \leftarrow j$.

\STATE \textbf{Step 2: Subproblem}
\STATE Solve \eqref{eq:sp} with $\mathbf{v}_j^*$. Get worst-case scenario $\mathbf{u}_j^*$ and recourse cost $Q(\mathbf{v}_j^*)$.
\STATE Update global upper bound $\overline{U} \leftarrow \min\{\overline{U}, \mathbf{c}^\top \mathbf{v}^*_j+Q(\mathbf{v}^*_j)\}$.

\STATE \textbf{Step 3: Optimality Test \& Backtracking}
\IF{ $(\overline{U} - L_\ell) / \overline{U} < \varepsilon$ }
    \STATE Terminate. Return $\mathbf{v}_j^*$.
\ELSIF{ $(\overline{U} - O_j) / \overline{U} < \tilde{\varepsilon}$ } 
    \STATE \textbf{Exploitation:} Set $j \leftarrow \ell$ and $\overline{L} \leftarrow L_\ell$. Tighten tolerance $\varepsilon^{\mathrm{MP}}_{j'} \leftarrow \alpha \varepsilon^{\mathrm{MP}}_{j'}, \forall j' \ge \ell $.
\ELSE
    \STATE \textbf{Exploration:} Update $\overline{L} \leftarrow O_j$. 
    \STATE \textbf{Scenario Set Enlargement:}
    Add $\mathbf{u}_j^*$ to $\mathcal{O}$. Add constraints \eqref{eq:compact_second_ineq} to \eqref{eq:mp}. $j \leftarrow j + 1$.
\ENDIF

\STATE Go to \textbf{Step 1}.
\end{algorithmic}
\end{algorithm}

\section{Numerical Experiment} \label{sec: numerical}
\begin{figure}[t]
    \centering
    \includegraphics[width=0.75\linewidth]{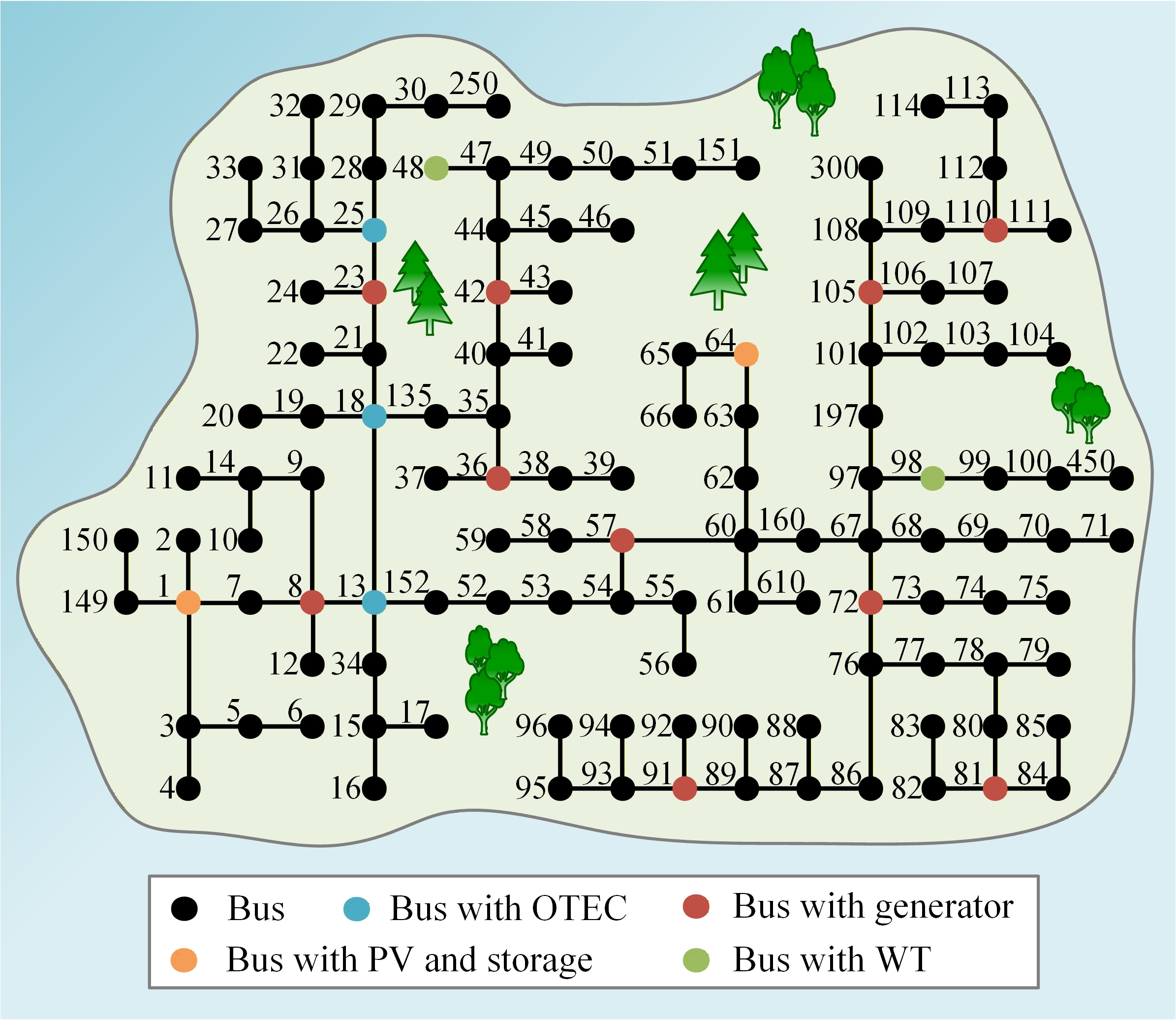}
    \caption{Modified IEEE 123-node test feeder layout.}
    \label{fig:IEEE 123}
\end{figure}

The proposed model is validated on a modified IEEE 123-node test feeder, which is adapted to operate as an island microgrid, as shown in Fig. \ref{fig:IEEE 123}.
We consider an island located in the South China Sea, with the sea surface temperature derived from historical data in autumn~\cite{datashareclub}.
We obtain the electricity demand data from~\cite{liu2018economic} and freshwater demand data from~\cite{mostafavi2018residential}.
The unit OTEC cost is 0.196 \textyen/kW~\cite{calvo2025ocean}, the unit energy storage operation cost is 0.0293 \textyen/kW~\cite{wu201628}, and the unit electrical demand response cost is 1.38 \textyen/kW~\cite{rana2023}.
Technical parameters for open-cycle OTEC stem from Penny et al. \cite{penny1984open}.
The default simulation parameters are configured as follows.
In the uncertainty set, the maximum relative deviation $\delta_i$ is set to 20\%. 
The budget parameter $\Gamma$ is set to 5. 
For the iCCG algorithmic configuration, the global optimality tolerance $\varepsilon$ is set to 1.5\%, and the inexact relative gap tolerance $\tilde{\varepsilon}$ is 1\%.
The optimality tolerance for the master problem $\varepsilon^{\mathrm{MP}}_0$ is initialized at 2\% with a contraction factor $\alpha$ of 0.5.
All experiments were conducted on a workstation equipped with an Intel Core Ultra 7 155H processor and 32 GB of RAM. 
The optimization is implemented in Python 3.13 using Gurobi 12.0.1 as the solver.
The maximum computation time is set to 20,000 seconds.
\begin{figure}[t]
  \centering
  \begin{subfigure}[b]{0.45\linewidth}
    \includegraphics[width=\linewidth]{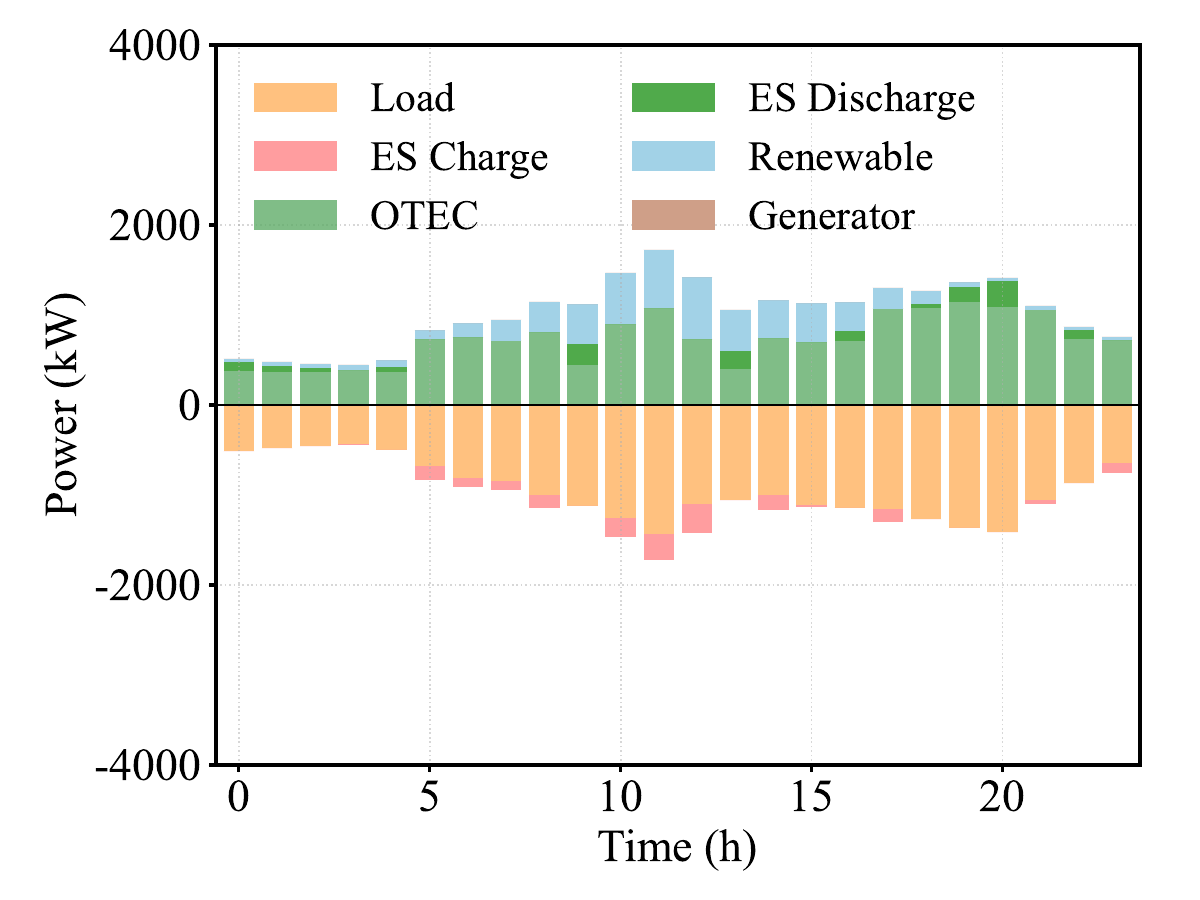}
    \caption{Electricity energy flow}
    \label{fig:sub1}
  \end{subfigure}
  \begin{subfigure}[b]{0.45\linewidth}
    \includegraphics[width=\linewidth]{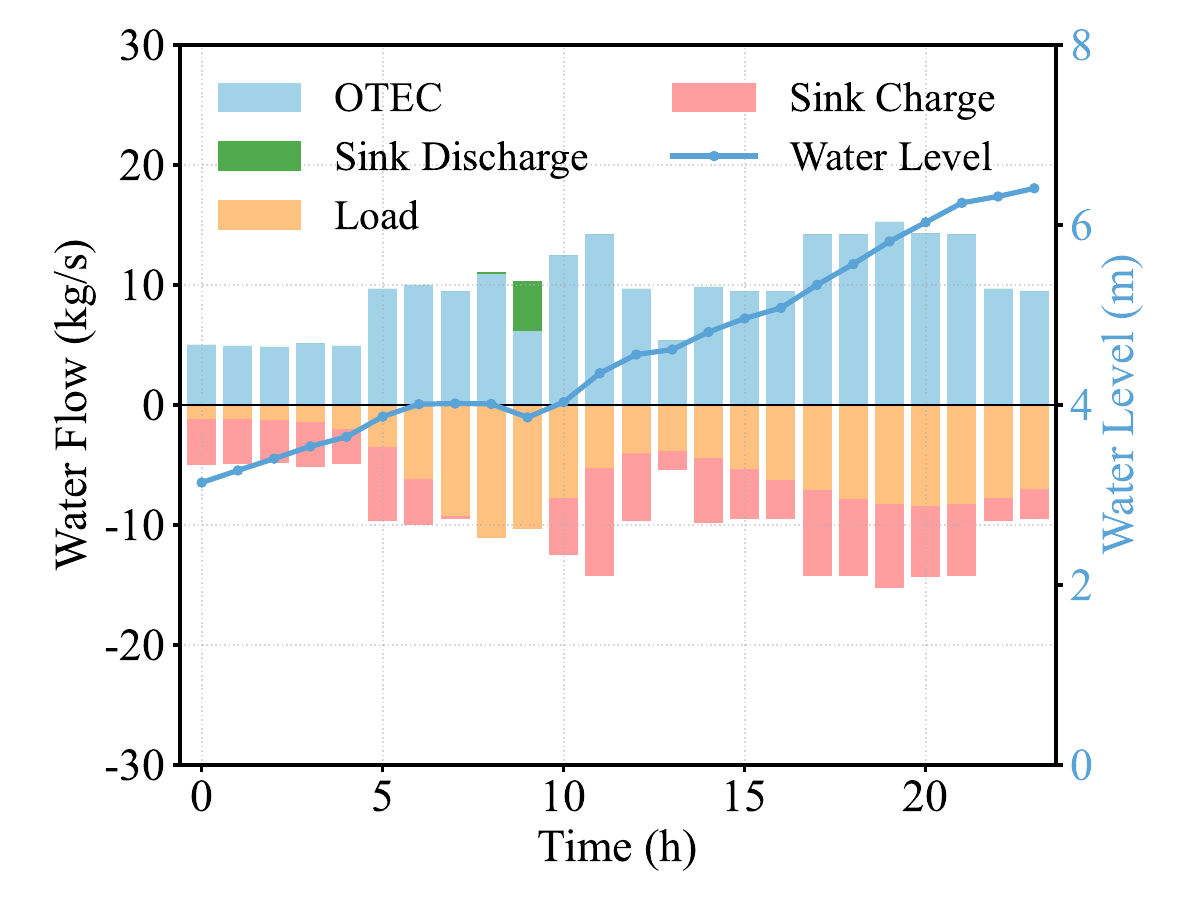}
    \caption{Freshwater mass flow}
    \label{fig:sub2}
  \end{subfigure}
  \caption{Operation results of the default scenario.}
  \label{fig:flow}
\end{figure}

\subsection{Optimal Scheduling Results under Default Scenario}
The optimal operation of the microgrid was simulated over a 24-hour horizon under the default scenario and parameters.
The resulting dispatch schedules for electrical power and freshwater are illustrated in Fig. \ref{fig:flow}.
Driven by the flexibility and lower costs compared to generators, OTEC units function as the main power source for most of the day, ensuring a reliable electricity supply. 
This indicates that OTEC systems can fully substitute fuel-based generators; and hence, compared to the volatile PVs and WTs, OTECs can offer a more dependable zero-carbon solution for island microgrids.
Furthermore, due to the inherent coupling between power generation and steam condensation in the turbine, the high electricity output of the OTEC units results in excessive freshwater production.
Consequently, surplus freshwater beyond residential demand is diverted to the water sink.

\begin{figure}[ht]
  \centering
  \begin{subfigure}[b]{0.48\linewidth}
    \includegraphics[width=\linewidth]{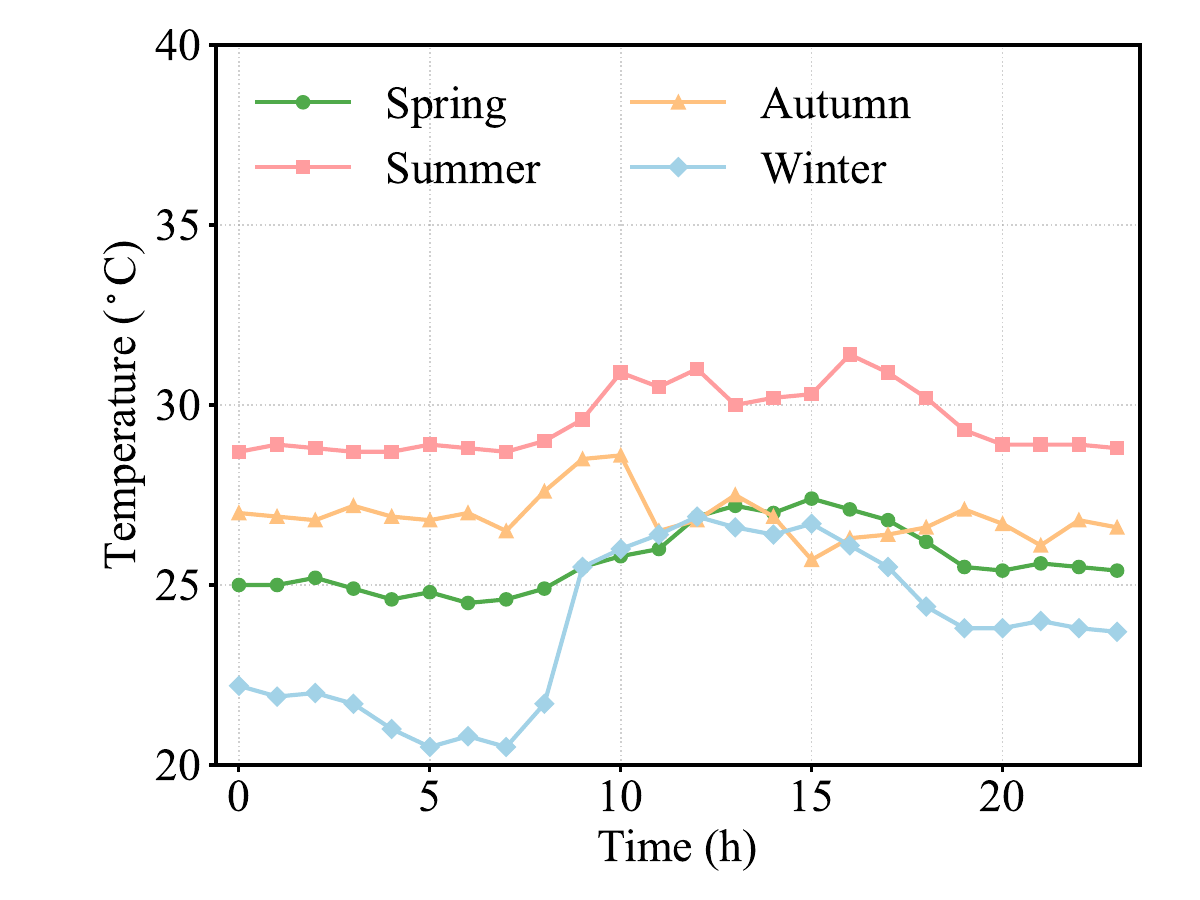}
    \caption{Temperature}
    \label{fig:season temp}
  \end{subfigure}
  \begin{subfigure}[b]{0.48\linewidth}
    \includegraphics[width=\linewidth]{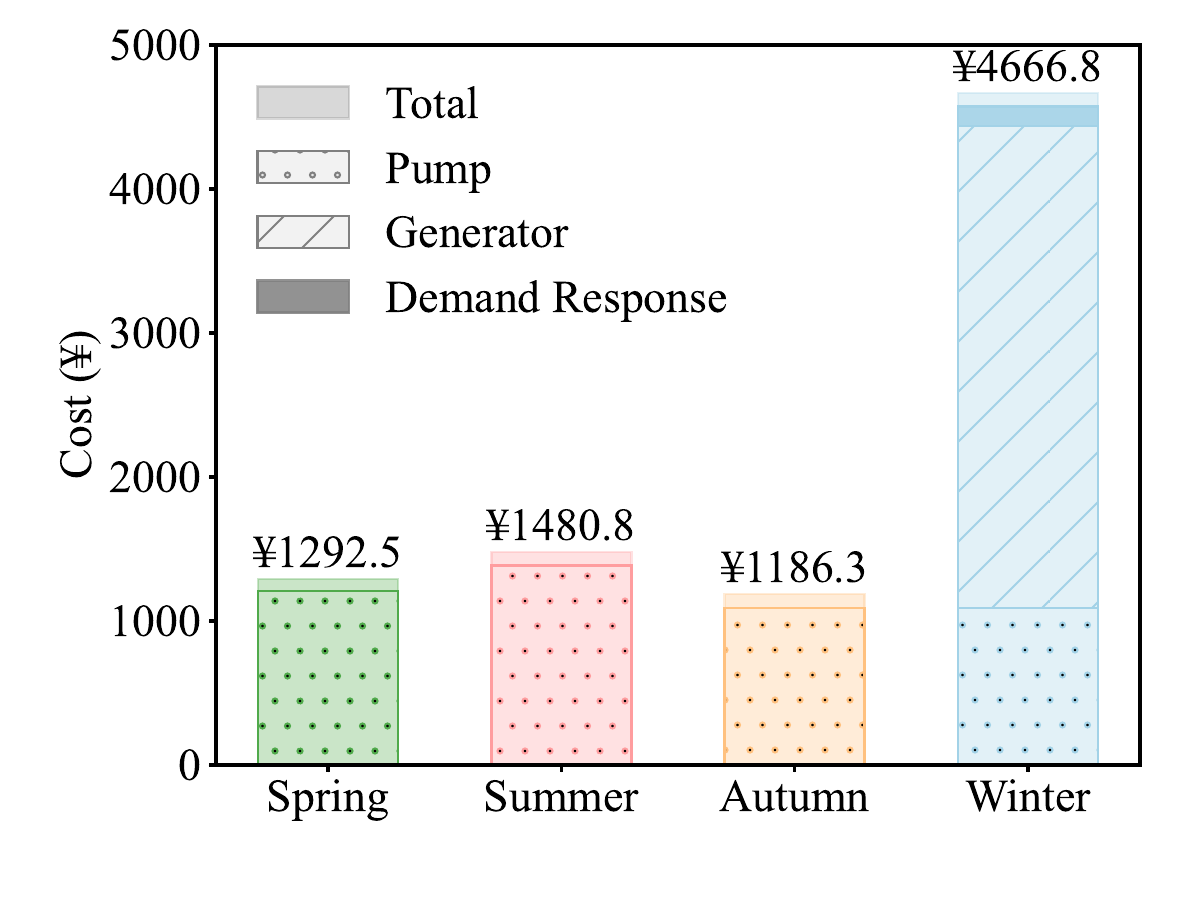}
    \caption{Operational cost}
    \label{fig:season cost}
  \end{subfigure}
  \caption{Seasonal temperatures and operational costs.}
\end{figure}

\subsection{Impact of Inlet Temperature on OTEC Power Output}

\begin{figure*}[ht]
    \centering
    \includegraphics[width=0.9\linewidth]{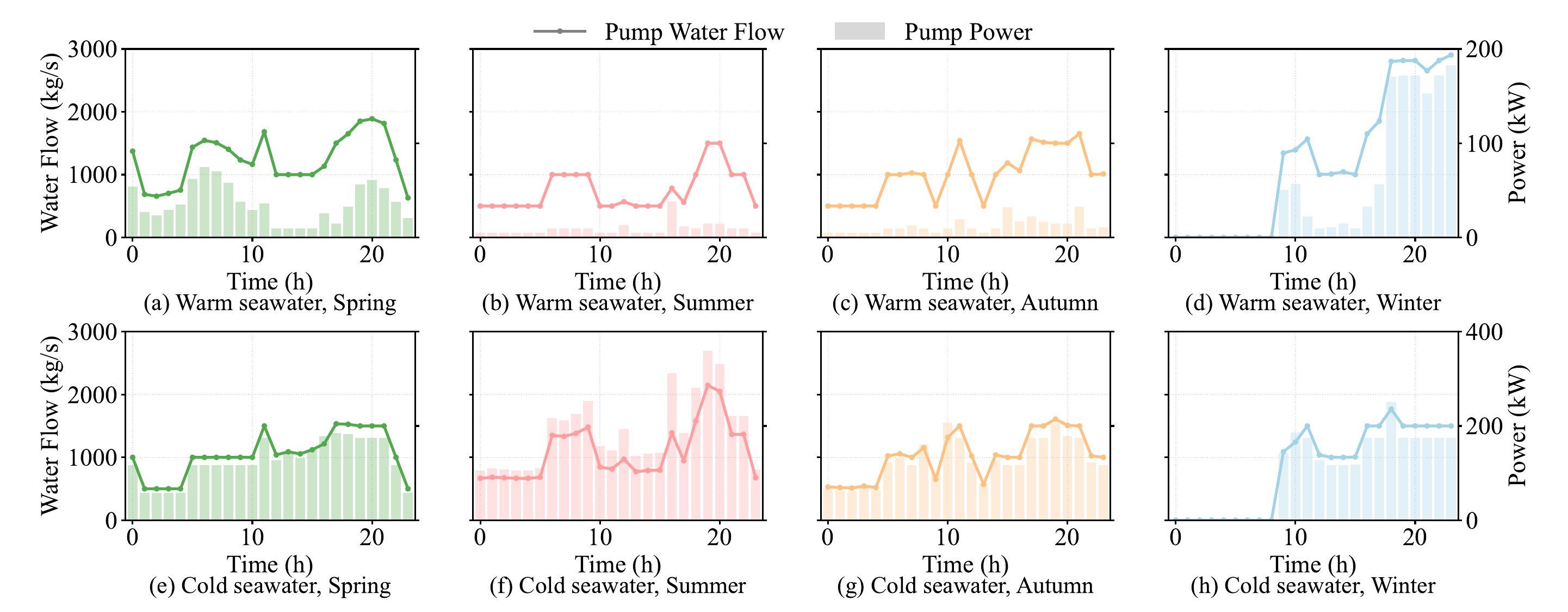}
    \caption{Seasonal operation details of OTEC units. The top row (a)--(d) shows the warm seawater flow rate and pump power, while the bottom row (e)--(h) shows the corresponding data for cold seawater.}
    \label{fig:otec_combined}
\end{figure*}

We also conducted simulations under temperatures of typical days in spring, summer, and winter, as illustrated in Fig.~\ref{fig:season temp}.
The overall operational costs follow the trend winter $>$ summer $>$ spring $>$ autumn, as shown in Fig. \ref{fig:season cost}.
The seasonal variations in flow rates and pump power consumption for OTEC operations are illustrated in Fig. \ref{fig:otec_combined}.

The trend across winter, spring, and autumn is consistent with the theoretical expectation that higher temperatures always yield lower costs.
In winter, due to the low inlet temperature, OTEC units frequently work in low-temperature conditions where neither electricity nor freshwater is produced.
Consequently, the system must aggressively increase the seawater flow rate during normal operation intervals (see Fig.~\ref{fig:otec_combined}d  and Fig.~\ref{fig:otec_combined}h),
and rely on conventional generators and demand response to make up for the shortfall in both power and freshwater supply, making the cost nearly four times that of other seasons.

Unexpectedly, despite higher temperatures, the operational cost in summer surpasses that in spring and autumn. 
This is primarily attributed to the minimum flow rate constraint of the warm seawater pump.
Although under high summer temperature, the evaporator could theoretically generate enough steam with much less warm seawater,
the constraint forces a minimum intake of at least 500 kg/s of warm seawater (Fig.~\ref{fig:otec_combined}b), resulting in excessive steam generation.
To fully condense the steam and maintain the thermodynamic balance, the system is forced to increase the cold seawater flow rate (Fig.~\ref{fig:otec_combined}f), causing higher power demand and pumping costs.

\subsection{Impact of Demand on Total Costs}
\begin{figure}[ht]
  \centering
  \begin{subfigure}[b]{0.46\linewidth}
    \includegraphics[width=\linewidth]{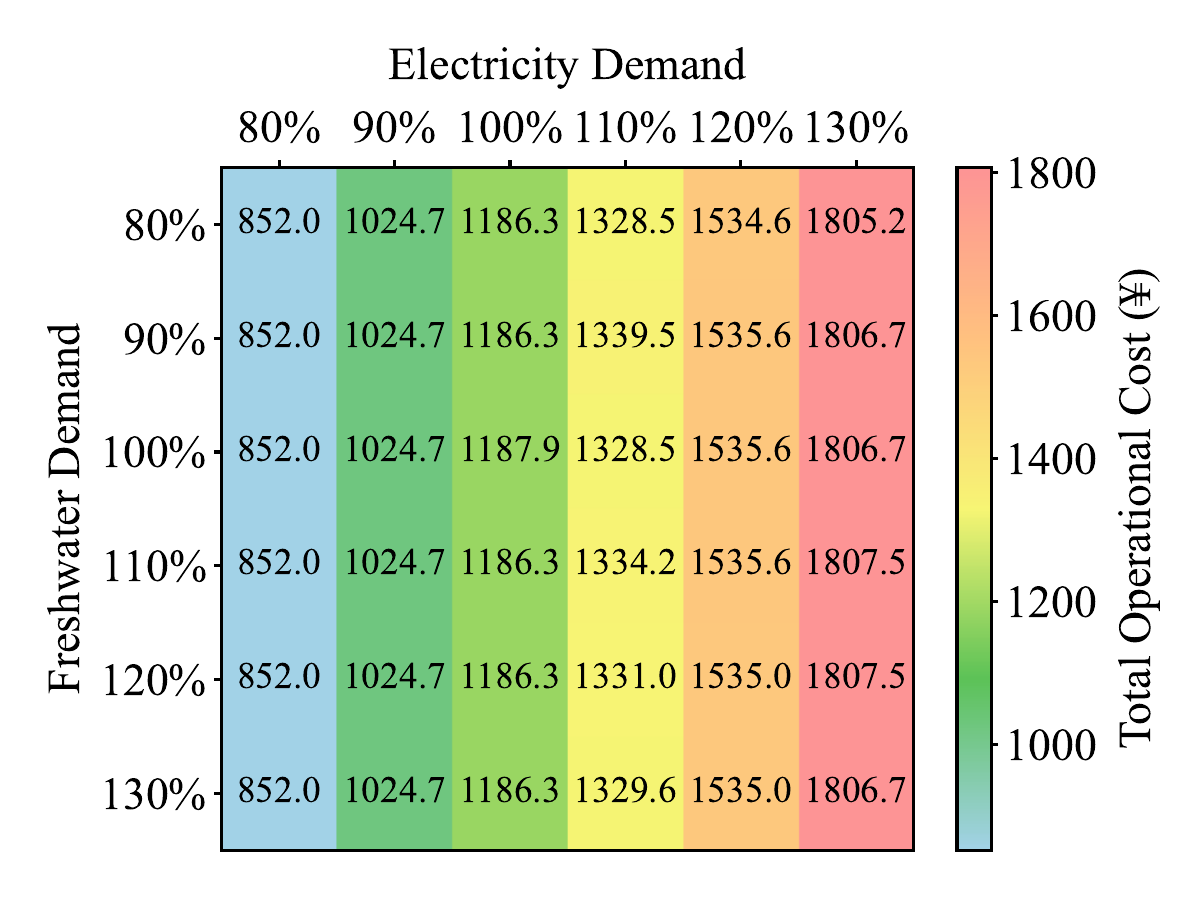}
    \caption{$\bar{h}=20\mathrm{m}$}
    \label{fig:costsub1}
  \end{subfigure}
  \begin{subfigure}[b]{0.46\linewidth}
    \includegraphics[width=\linewidth]{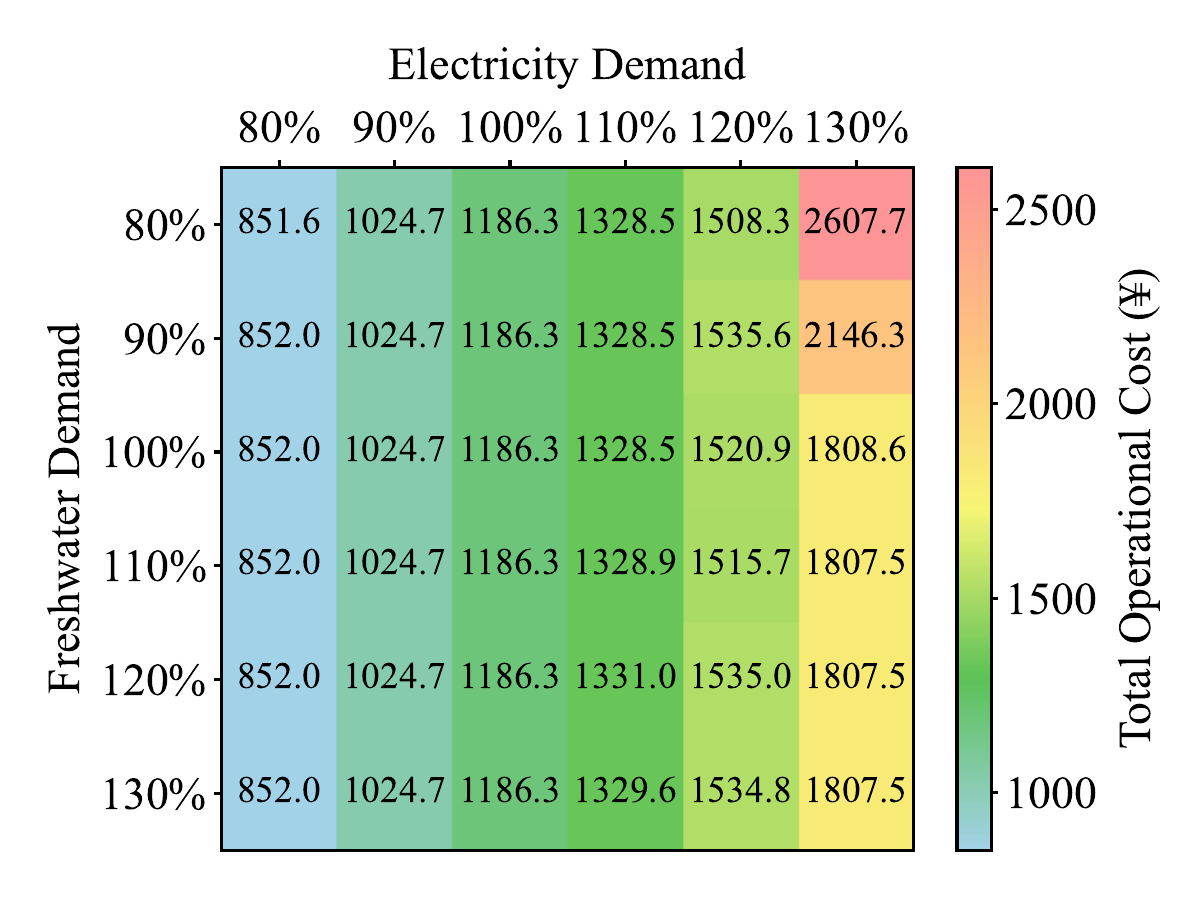}
    \caption{$\bar{h}=10\mathrm{m}$}
    \label{fig:costsub2}
  \end{subfigure}
  \caption{Total costs under different demand conditions.}
  \label{fig:totalcost}
\end{figure}

To evaluate the economic performance of the proposed system under varying load conditions, we scaled the baseline freshwater and electricity demand profiles from 80\% to 130\%, while keeping all other parameters at their default values.
As illustrated in Fig. \ref{fig:costsub1}, the cost gradient is steeper along the electricity axis compared to the water axis, reflecting a higher marginal cost for power generation.
This is because the freshwater production capacity of OTEC exceeds its power generation capacity.
Under the baseline scenario, satisfying the electrical load naturally yields a freshwater output that surpasses the actual water demand.
This provides a buffer against increases in freshwater demand.
In contrast, when electricity demand rises, OTEC units must ramp up power output, which increases pumping and operational costs.

\begin{figure}[ht]
  \centering
  \begin{subfigure}[b]{0.46\linewidth}
    \includegraphics[width=\linewidth]{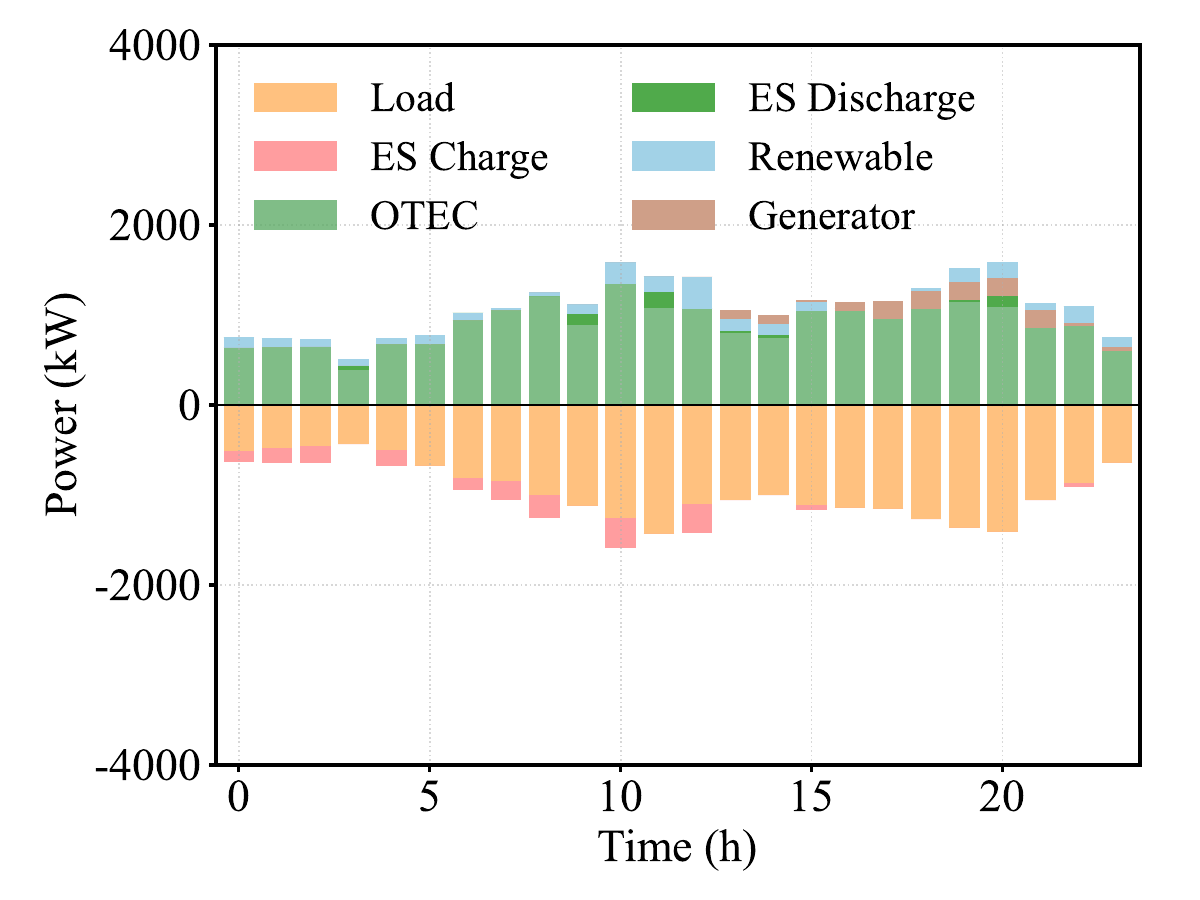}
    \caption{Electricity energy flow}
    \label{fig:balsub1}
  \end{subfigure}
  \begin{subfigure}[b]{0.46\linewidth}
    \includegraphics[width=\linewidth]{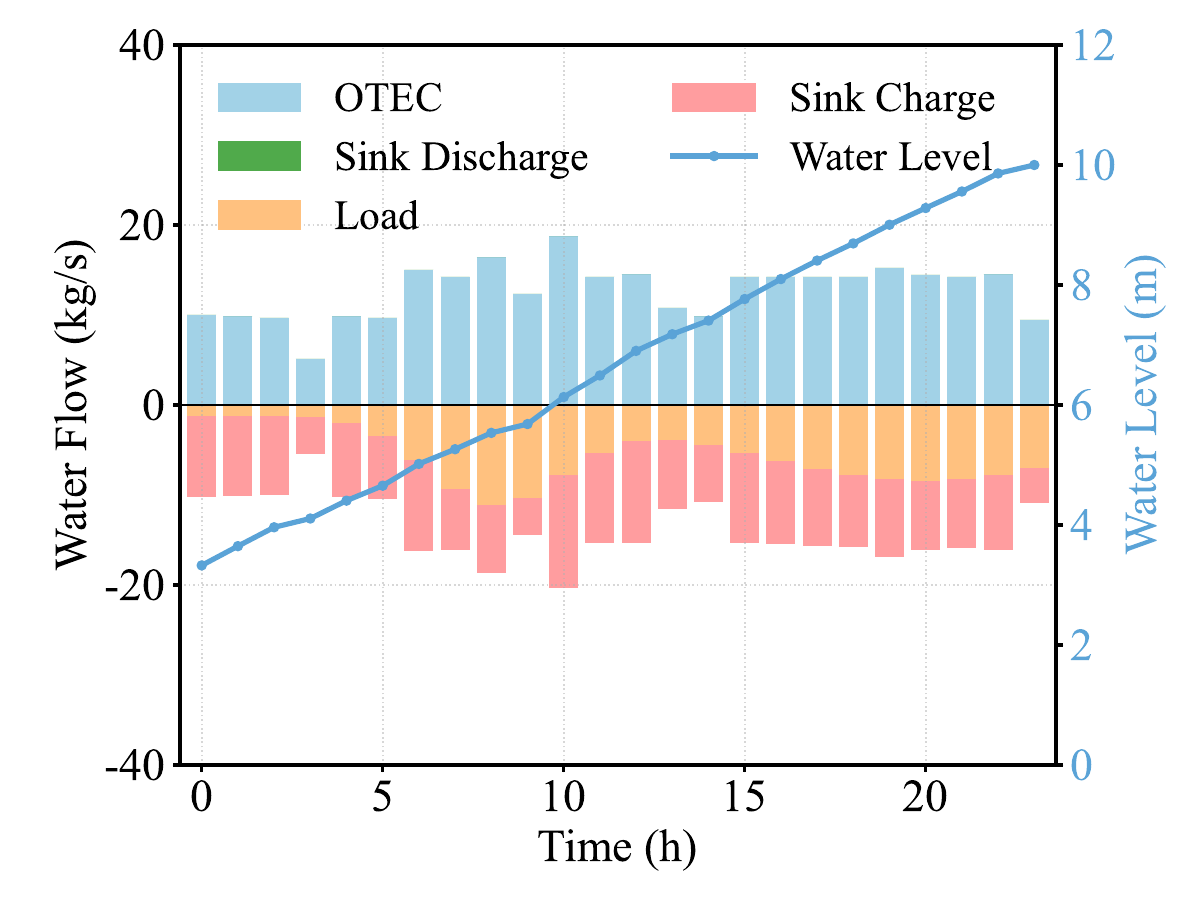}
    \caption{Freshwater mass flow}
    \label{fig:balsub2}
  \end{subfigure}
  \caption{Operation results of $\bar{h}=10\mathrm{m}$, 80\% water demand, and 130\% electricity demand.}
  \label{fig:bal2}
\end{figure}

In addition to the default simulation scenario with a maximum sink level of $\bar{h}=20 \mathrm{m}$, we repeated the sensitivity analysis for $\bar{h}=10\mathrm{m}$.
The results are presented in Fig.~\ref{fig:costsub2} and Fig.~\ref{fig:bal2}.
In this configuration, although OTEC units are dispatched to meet higher electrical loads, their output is constrained by the reduced capacity of the water sink.
This is in accordance with Fig. \ref{fig:balsub2}, where the sink water level has reached its maximum.
As a result, the system is forced to commit conventional generators to satisfy electricity demand, which brings additional costs.
This case demonstrates the necessity of coordinating both the water and electricity demand when scheduling the open-cycle OTEC.

\begin{figure}[!ht]
    \centering
    \includegraphics[width=0.75\linewidth]{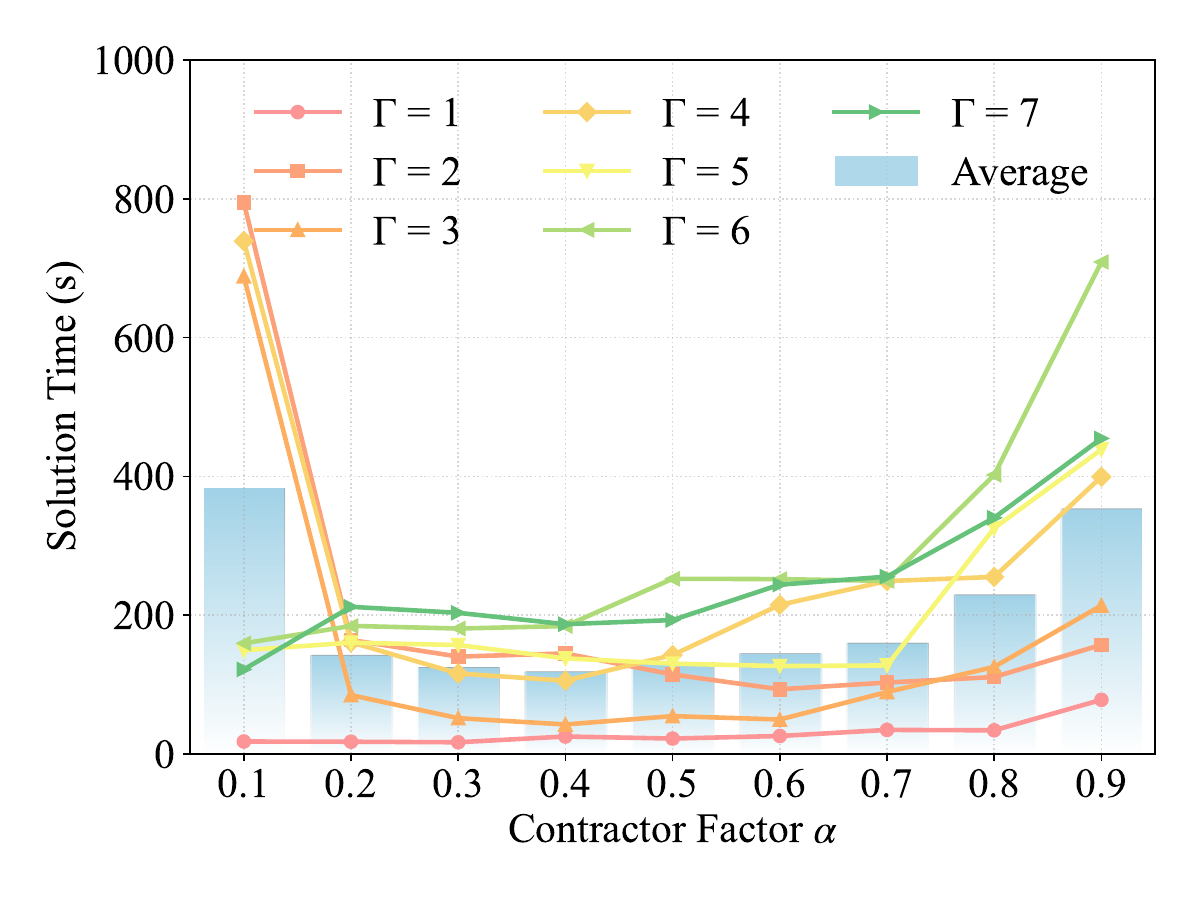}
    \caption{Solution time under different contraction factors.}
    \label{fig:alpha}
\end{figure}

\subsection{Computational Performance}
To validate the efficiency and robustness of the proposed iCCG algorithm, we conducted a systematic performance evaluation focusing on two key dimensions.

\subsubsection{Budget Parameter $\Gamma$}
We varied the uncertainty budget $\Gamma$ from 1 to 9 and evaluated the performance against the standard CCG under varying degrees of conservatism.
For a fair comparison, both algorithms were initialized with the same nominal scenario and terminated upon reaching a 1.5\% global optimality gap.
The results are summarized in Table \ref{tab:budget parameter}.
\begin{table}[!htbp]
  \centering
  \caption{Impact of Budget Parameter $\Gamma$ on Computational Performance}
  \label{tab:budget parameter}
  \begin{tabular}{ccccccc}
    \toprule
    $\Gamma$ & Algo. & Cost (\textyen) & Time (s) & MP Time (s) & Explor. & Exploit. \\
    \midrule
    \multirow{2}{*}{1} & CCG  & 1172.60 & 243.28 & 179.96 & 1 & -- \\
                       & iCCG & 1172.60 &  22.37 &   5.09 & 1 & 2 \\
    \midrule
    \multirow{2}{*}{2} & CCG  & 1179.36 & 899.60 & 455.14 & 2 & -- \\
                       & iCCG & 1179.36 & 114.70 &  12.76 & 2 & 1 \\
    \midrule
    \multirow{2}{*}{3} & CCG  & 1185.70 & 805.97 & 521.95 & 2 & -- \\
                       & iCCG & 1183.62 & 393.71 &   8.98 & 2 & 1 \\
    \midrule
    \multirow{2}{*}{4} & CCG  & 1185.00 & 1197.63 & 929.10 & 3 & -- \\
                       & iCCG & 1185.00 &  142.70 &  11.43 & 2 & 1 \\
    \midrule
    \multirow{2}{*}{5} & CCG  & 1186.27 &  897.26 & 551.69 & 2 & -- \\
                       & iCCG & 1187.86 &  130.29 &   7.33 & 2 & 1 \\
    \midrule
    \multirow{2}{*}{6} & CCG  & 1187.53 & 2050.76 & 572.38 & 2 & -- \\
                       & iCCG & 1187.53 &  252.58 &  17.63 & 3 & 2 \\
    \midrule
    \multirow{2}{*}{7} & CCG  & 1196.12 & 1548.29 & 595.29 & 2 & -- \\
                       & iCCG & 1189.53 &  193.11 &   7.91 & 2 & 1 \\
    \midrule
    \multirow{2}{*}{8} & CCG  & 1198.17 & 1011.14 & 597.60 & 2 & -- \\
                       & iCCG & 1198.17 & 13452.29 & 657.06 & 3 & 2 \\
    \midrule
    \multirow{2}{*}{9} & CCG  & -- & $>20000$ & -- & -- & -- \\
                       & iCCG & -- & $>20000$ & -- & -- & -- \\
    \bottomrule
  \end{tabular}
\end{table}

As expected, the total costs for CCG and iCCG generally increase with $\Gamma$.
At lower uncertainty budgets $\Gamma\le 7$, iCCG demonstrates superior computational speed compared to standard CCG. 
This advantage stems from iCCG's mechanism, which enables rapid exploration through inexact solving and verifies strict optimality during the exploitation steps when necessary.
Consequently, iCCG significantly reduces the solution time of \eqref{eq:mp} as well as the total runtime.
At a high uncertainty budget $\Gamma=8$, the standard CCG outperforms iCCG in speed.
This is perhaps because iCCG naturally requires additional exploitation iterations to verify the optimality, and the more challenging subproblem will add more computational burden to every iteration, resulting in a notable increase in total solution time.
When $\Gamma=9$, there can be up to $1.3$ trillion scenarios in $\mathcal{U}$, bringing enormous computational burdens, and neither of the two algorithms can solve the problem within the 20,000-second time limit. 
Hence, we focus on scenarios with $\Gamma \leq 7$ in the following numerical studies.

\subsubsection{Contraction Factor $\alpha$}
We performed a sensitivity analysis on the contraction factor $\alpha$ from 0.1 to 0.9 and the budget parameter $\Gamma$ from 1 to 7, as shown in Fig. \ref{fig:alpha}.
The average solution times under different budget parameters $\Gamma$ are generally lower when $\alpha \in [0.2, 0.7]$, but become higher at $\alpha=0.1$ or $\alpha=0.9$.
This is because a very small $\alpha$ imposes an excessively tight optimality tolerance for the master problem $\varepsilon^{MP}_j$ during the exploitation step, while a very large $\alpha$ slows the contraction of $\varepsilon^{MP}_j$ and requires more exploitation steps to meet the global optimality tolerance, both leading to longer solution time.
Hence, a moderate $\alpha$ is recommended.

\section{Conclusion} \label{sec: conclusion}
This paper develops a model of the thermodynamic process and operational framework of open-cycle OTEC and embeds it in an island microgrid scheduling problem. 
To address the intermittency of RESs, we formulate a two-stage robust operation model with a budget uncertainty set and solve it using an iCCG algorithm. 
Numerical studies indicate that open-cycle OTEC can fully replace conventional generators while offering higher reliability than undispatchable renewables. 
Microgrid operating costs exhibit strong seasonal dependence: winter conditions drive frequent low-temperature operation and reduced OTEC output, nearly quadrupling total costs relative to other seasons, whereas summer conditions increase steam production and the required cold-seawater flow, yielding unexpectedly higher costs than in spring and autumn. 
Computationally, the proposed iCCG improves efficiency, reducing runtime by up to 90\% compared with standard CCG. 
Overall, open-cycle OTEC emerges as a reliable, zero-carbon option for coupled energy--water supply on remote islands, and the proposed modeling and solution framework can be extended to other coastal and offshore microgrids, e.g., drilling platforms and marine farming facilities.

\appendices
\section*{Appendix}
Due to the limited space, we provide the Appendix online at \url{https://github.com/Xiaoyu-Fu/Appendices.git}.

\normalem
\bibliographystyle{IEEEtran}
\bibliography{IEEEabrv,reference}
\end{document}